\documentclass[12pt]{article}

\usepackage{amsthm}
\usepackage{mathtools}
\usepackage{amsmath}
\usepackage{amssymb}
\DeclareMathAlphabet{\mathbbold}{U}{bbold}{m}{n}
\newcommand*{\boldone}{\mathbbold{1}}
\allowdisplaybreaks

\usepackage[T1]{fontenc}
\usepackage[utf8]{inputenc}
\usepackage{lmodern}
\usepackage[protrusion=true, expansion=true]{microtype}
\usepackage{setspace}

\usepackage[english]{babel}
\usepackage{csquotes}

\usepackage[a4paper, margin=1in]{geometry}

\usepackage{booktabs}
\usepackage[hang, small, labelfont=bf, textfont=it, up]{caption}
\usepackage{enumerate}
\usepackage{graphicx}
\usepackage{threeparttable}

\newtheorem{remark}{Remark}
\newtheorem{assumption}{Assumption}
\newtheorem{example}{Example}

\newtheorem{theorem}{Theorem}

\usepackage{abstract}

\usepackage[usenames, dvipsnames]{xcolor}

\usepackage[colorlinks=true, linkcolor=blue, urlcolor=blue, citecolor=blue]{hyperref}

\usepackage[authordate, backend=biber]{biblatex-chicago}
\usepackage{titling}

\title{
{\Large\bfseries
Debiased Fixed Effects Estimation of Binary Logit Models with Three-Dimensional Panel Data
}\thanks{Amrei Stammann is thankful for financial support from Mercator Research Center Ruhr (Mercur) within the project ``Digitale Daten in der sozial- und wirtschaftswissenschaftlichen Forschung''. She thanks Daniel Czarnowske, Jannis Kück, Cavit Pakel, Martin Schumann, and Joschka Wanner
for helpful comments and discussions. The estimators from this paper are implemented in the R-package \href{https://cran.r-project.org/package=alpaca}{alpaca}.}\\[0.5em]
}
\author{
Amrei Stammann
\thanks{
Ruhr-Universität Bochum, Universitätsstr. 150, 44801 Bochum, Germany and Research Data Center at RWI-Leibniz Institute for Economic Research, Hohenzollernstr. 1--3, 45128 Essen, Germany; e-mail: \texttt{\href{mailto:amrei.stammann@rub.de}{amrei.stammann@rub.de}}
}
}
\date{\small\today}

\DeclareMathOperator{\argmin}{\arg\,\min\;}
\DeclareMathOperator{\argmax}{\arg\,\max\;}
\DeclareMathOperator{\diag}{\text{diag}}
\DeclareMathOperator{\eye}{\mathbb{I}}
\DeclareMathOperator{\iid}{\text{iid.}\;}
\DeclareMathOperator{\ind}{\boldone}
\DeclareMathOperator{\N}{\mathcal{N}}
\DeclareMathOperator{\U}{\mathcal{U}}

\newcommand{\norm}[1]{\lVert #1 \rVert}
\newcommand{\bignorm}[1]{\big\lVert #1 \big\rVert}

\newcommand{\EX}[1]{\mathbb{E}\left[ #1 \right]}
\newcommand{\CEX}[1]{\mathbb{E}\left[#1 \, \middle\vert \, \mathcal{F}\right]}
\newcommand{\LEX}[1]{\overline{\mathbb{E}}\left[ #1 \right]}
\theoremstyle{definition}

\newtheorem{lemma}{Lemma}

\begin{document}

\maketitle

\thispagestyle{empty}
\renewcommand{\abstractname}{\vspace{-5em}}
\begin{abstract}
    Naive maximum likelihood estimation of binary logit models with fixed effects leads to unreliable inference due to the incidental parameter problem. We study the case of three-dimensional panel data, where the model includes three sets of additive and overlapping unobserved effects. This encompasses models for network panel data, where senders and receivers maintain bilateral relationships over time, and fixed effects account for unobserved heterogeneity at the sender-time, receiver-time, and sender-receiver levels. In an asymptotic framework, where all three panel dimensions grow large at constant relative rates, we characterize the leading bias of the naive estimator. The inference problem we identify is particularly severe, as it is not possible to balance the order of the bias and the standard deviation. As a consequence, the naive estimator has a degenerating asymptotic distribution, which exacerbates the inference problem relative to other fixed effects estimators studied in the literature. To resolve the inference problem, we derive explicit expressions to debias the fixed effects estimator.
	\vfill
\end{abstract}

\clearpage
\onehalfspacing

\setcounter{page}{1}
\section{Introduction}

%\textbf{Part I: What is the general topic and why is it important?}\\
Even after more than 75 years since its discovery by \textcite{ns1948}, the incidental parameter problem, originally a specific inconsistency problem that occurs in many fixed effects estimators for nonlinear models, remains a highly studied topic in panel data econometrics. Most work on the incidental parameter problem focuses on \textit{classical} panel data sets, i.e. \textit{two-dimensional} panels, where cross-sectional units $N$ are observed over several time periods $T$. Here, the incidental parameter problem arises under asymptotics where $N$ tends to infinity, while $T$ is fixed. One approach to tackle the incidental parameter problem is to rely on a different asymptotic framework, usually $N, T \rightarrow \infty$. This fixes the inconsistency problem but introduces an asymptotic bias problem. If this asymptotic bias problem is not properly addressed, e.g.\ by applying suitable bias corrections, the fixed effects estimator becomes unreliable for drawing inferences. In this paper, we focus on bias correction approaches developed under large-$T$ asymptotics. However, it is worth mentioning another important strand of literature that focuses on developing fixed-$T$ consistent estimators. These estimators typically rely on eliminating the unobserved effects from the model by differencing or conditioning on sufficient statistics.\footnote{Examples for binary logit models are, among others, \textcite{r1960}, \textcite{a1970}, \textcite{c1980}, \textcite{hk2000}, or \textcite{hw2022}.}

\textit{Multi-dimensional} panel data is becoming increasingly common in empirical research, as data becomes more granular. One example of a \textit{three-dimensional} panel is data on bilateral network activities observed over time, which is often used in international trade research. More specifically, researchers may study trade flows between $I$ exporting countries and $J$ importing countries over $T$ years. This multi-dimensionality allows to control for richer sources of unobserved heterogeneity, leading to model specifications with \textit{multi-way} fixed effects. For example, researchers in international trade often control for unobserved heterogeneity at the exporter-time, importer-time, and importer-exporter levels in their empirical analyses.\footnote{Controlling for unobserved heterogeneity at the exporter-time, importer-time, and importer-exporter levels is, among others, recommended in \textcite{hm2014}.} However, the asymptotic properties of the corresponding fixed effects estimators are largely unknown, with a few exceptions.

%\textbf{Part II: What is it that I do?}\\
In this paper, we derive the asymptotic properties of fixed effects estimators for static binary logit models for three-dimensional panels, where the three sets of unobserved effects enter additively into the linear index as $\alpha_{it} + \gamma_{jt} + \rho_{ij}$. Under asymptotics, where all panel dimensions grow large and $I \sim J \sim T$, we use expansions to characterize the leading bias term and suggest an appropriate bias correction. We show that the order of the bias is $1 / I + 1 / J + 1 / T$, confirming a conjecture of \textcite{fw2018} that has not been proven yet.\footnote{The conjecture is based on a heuristic formula developed by \textcite{fw2018} as part of a review of recent advances in fixed effects estimation.} Moreover, we confirm the correctness of the expressions conjectured and proposed by \textcite{hsw2020} for bias-corrected estimators. 

%\textbf{Part III: What do I find?}\\
Our main finding, which distinguishes our case from all other cases studied in the bias correction literature, is that the inference problem is more severe. In most cases, the order of the bias and standard deviation of fixed effects estimators are the same, allowing authors to derive \textit{non-degenerate} asymptotic distributions for the uncorrected estimators, which are centered around distorted expected values. Bias corrections then center the asymptotic distributions correctly around zero. Our inference problem is particularly severe because the leading bias of our fixed effects estimator, $1 / I + 1 / J + 1 / T$, is of a higher order than its standard deviation, $1 / \sqrt{IJT}$, which leads to a \textit{degenerating} asymptotic distribution. Therefore, developing a debiased estimator is particularly important. In simulation experiments we confirm the severity of the inference problem without bias correction, as confidence intervals constructed around the uncorrected estimator almost never cover the true model parameters, even in large samples. Our proposed bias correction is effective in improving the inferential accuracy of the fixed effects estimator. An empirical example from international trade shows that debiased estimates can differ substantially from uncorrected estimates in real-world applications. Thus, our findings have important implications for empirical researchers, highlighting the need for bias correction to obtain reliable inference. We expect our results to generalize to other link functions and dynamic models.  To simplify the analysis, we study the asymptotic properties of fixed effects estimators using the
binary logit model as an example, following arguments of \textcite{cfw2020}.\footnote{The simplification mainly comes from the fact that high-order derivatives of the log-likelihood function no longer depend on the outcome variable.} We plan to generalize our asymptotic analysis to (dynamic) nonlinear models with concave objective functions in the future.

%\textbf{Part IV: How does it relate to what we already know?}\\
A large part of the previous large-$T$ literature has focused on providing solutions for classical panel data models with individual effects, see among others, \textcite{hk2002}, \textcite{l2002}, \textcite{w2002}, \textcite{s2003}, \textcite{hn2004}, \textcite{c2007}, \textcite{ab2009}, \textcite{bh2009}, \textcite{f2009}, \textcite{hk2011}, \textcite{dj2015}, \textcite{ks2016}, \textcite{p2019}, \textcite{sst2021}, \textcite{hj2022}, or \textcite{s2023}. These proposed solutions differ in various ways, including the assumptions they make, the methods used to derive them, and the types of corrections they propose. We refer the reader to \textcite{ah2007} and \textcite{fw2018} for comprehensive reviews of this strand of literature. \textcite{fw2016} advance the literature by developing solutions for nonlinear panel data models that can account for both individual and time effects.\footnote{The authors' analysis is not limited to panels with time as the second panel dimension. It can also be applied to other two-dimensional panels, such as panels where the second panel dimension is another cross-section, such as countries or industries. For example, their analysis could be used to study a cross-section of bilateral trade flows between countries or a cross-section of patent citations between industries.} This is a major contribution, as accounting for both types of effects is challenging. We will discuss their contribution in more detail in Section \ref{sec:asymptotic_theory}, as it is essential for the derivation of our results. \textcite{jo2019} also studied nonlinear panel data models with individual and time effects. However, instead of correcting the bias in the asymptotic distribution, as \textcite{fw2016}, they proposed a likelihood correction approach.\footnote{Recently, \textcite{lms2023} presented a related approach to \textcite{jo2019} and additionally proved the asymptotic properties of the corrected likelihood and the test statistics of the trinity tests of maximum likelihood estimation (Wald, Lagrange-multiplier, and Likelihood-ratio test).} Contemporaneously with the development of \textcite{fw2016}'s bias correction, \textcite{c2017} extended the conditional logit estimator of \textcite{r1960} and \textcite{c1980} to handle two-way fixed effects.\footnote{She applied her estimator to a cross-sectional model of bilateral export probability. Her estimator requires both panel dimensions to grow large, as proven later by \textcite{j2018}, who derived its asymptotic properties. Because her approach is already computationally demanding in bilateral cross-sections, applying it to bilateral panels or extending it to three-way fixed effects may not be feasible. Moreover, it is not possible to generalize her approach to other nonlinear models and weakly exogenous regressors.} There is only little research on fixed effects estimators for multi-dimensional panel models with multiple unobserved effects, although multi-dimensional panels are increasingly common in empirical studies. \textcite{wz2021} is the only other paper apart from ours that has theoretically analyzed the properties of a fixed effects estimator for a three-dimensional panel model with three additive and overlapping unobserved effects. In particular, they studied the properties of the fixed effects Pseudo-Poisson estimator for the gravity model, the workhorse model of international trade.\footnote{\textcite{yh2023} theoretically analyzed an estimator for an alternative gravity model with three multiplicative, instead of additive, and overlapping unobserved effects by extending the generalized method of moments (GMM) estimation strategy proposed by \textcite{j2017}.} Contrary to us, \textcite{wz2021} can exploit a unique property of the Poisson model to eliminate $\rho_{ij}$ from the linear index. This essentially reduces the problem to analyzing a two-way fixed effects model, which allows them to rely on the asymptotic analysis of \textcite{fw2016}. As a consequence, they only require $I$ and $J$ to grow to infinity, while we require all three panel dimensions to grow.

The rest of the paper is organized as follows. Section \ref{sec:model_and_estimator} introduces the model and the fixed effects estimator. Section \ref{sec:asymptotic_theory} presents the asymptotic theory and Section \ref{sec:differences} discusses the key differences from previous studies. Sections \ref{sec:simulation_experiments} and \ref{sec:empirical_example} report results of simulation experiments and an empirical example. Section \ref{sec:conclusion} concludes.

\section{Model and estimator}
\label{sec:model_and_estimator}

\subsection{Model}

We observe three-dimensional panel data $\{(y_{ijt}, x_{ijt}) \colon i \in \mathcal{I}, \, j \in \mathcal{J}, \, t \in \mathcal{T}\}$, where $y_{ijt}$ is a binary outcome variable, $x_{ijt}$ is a vector of strictly exogenous explanatory variables, $\mathcal{I} = \{1, \ldots, I\}$, $\mathcal{J} = \{1, \ldots, J\}$, and $\mathcal{T} = \{1, \ldots, T\}$. We consider the following semi-parametric binary logit model with additive unobserved effects: 
\begin{equation}
    \label{eq:model}
    y_{ijt} = \ind\{x_{ijt}^{\prime} \beta + \alpha_{it} + \gamma_{jt}+ \rho_{ij} \geq \epsilon_{ijt}\} \, , \quad \epsilon_{ijt} \mid x, \alpha, \gamma, \rho \sim F_{\epsilon} \, ,
\end{equation}
where $\ind\{\cdot\}$ is an indicator function, $x = (x_{111}, \ldots, x_{IJT})$, $\alpha = (\alpha_{11}, \ldots, \alpha_{IT} )$, $\gamma = (\gamma_{11}, \ldots, \gamma_{JT} )$, $\rho = (\rho_{11}, \ldots, \rho_{IJ})$, $\epsilon_{ijt}$ is an idiosyncratic error term, and $F_{\epsilon}$ is the logistic cumulative distribution function. Further, $\beta$ is a $K$-dimensional vector of model parameters, and $\alpha$, $\gamma$, and $\rho$ are $IT$-, $JT$-, and $IJ$-dimensional vectors of unobserved effects, respectively. We interpret the model as semi-parametric because we do not make assumptions about the relationship between the unobserved effects and the explanatory variables, nor do we make assumptions about the distributions of the unobserved effects.

Below we present two examples of three-dimensional panel data sets in which our model could be applied.

\begin{example}[Bipartite / Undirected Networks]\label{ex:bipartite}
    Panel data of firms often contains additional information that can be used to form a three-dimensional panel. This can be, amongst others, information about products, information about export destinations, or information about locations of subsidiaries. For example, \textcite{s2022} uses a firm-country-time panel to analyze whether firms increase their propensity to avoid taxes by moving to the same tax haven where another firm operating in the same industry already engages in tax avoidance.
\end{example}

\begin{example}[Bilateral / Directed Networks]\label{ex:bilateral}
    Panel data on bilateral relationships between countries is typically used in fields such as international trade or economics of migration. For example, in spirit of \textcite{hmr2008}, we could model the probability of country $i$ exporting to country $j$ at time $t$ as a function of trade cost variables. This could be interesting on its own or as a first step in a two-step Heckman-type sample selection procedure to explain bilateral trade flows.
\end{example}

\subsection{Fixed effects estimation}

We collect the incidental parameters in the vector $\phi = (\alpha, \gamma, \rho)$ and estimate them along with the model parameters $\beta$ by minimizing the following constrained negative log-likelihood function:
\begin{align}
    \label{eq:objective_function}
    L(\beta, \phi) =& \, - \frac{1}{\sqrt{NT}}  \sum_{i = 1}^{I} \sum_{j = 1}^{J} \sum_{t = 1}^{T} (y_{ijt} \log(\mu_{ijt}(\beta, \phi)) + (1 - y_{ijt}) \log(1 - \mu_{ijt}(\beta, \phi)))  \nonumber \\ 
    & \, + \frac{c_{1}}{2 \sqrt{NT}} \phi^{\prime} v v^{\prime} \phi \, ,
\end{align}
where $0 < c_{1} < \infty$, $\mu_{ijt}(\beta, \phi) = \mu(x_{ijt}^{\prime} \beta + w_{ijt}^{\prime} \phi)$, $\mu(z) = (1 + \exp(- z))^{- 1}$ is the logistic cumulative distribution function, and the matrix $w$ is a collection of $IT + JT + IJ$ indicator variables arising from ``dummy encoding'' the following interactions of the three panel indices: $i \times t$, $j \times t$, and $i \times j$. The matrix $v$ imposes constraints on the incidental parameters $\phi$ to ensure uniqueness of the solution of the optimization problem and therefore the second term in $L(\beta, \phi)$ acts as ``penalty'' term. Essentially, the penalty term prevents $w$ from being rank-deficient and therefore plays an important role in ensuring the invertibility of the incidental parameter Hessian,
\begin{equation}
    \label{eq:ip_hessian}
    \frac{\partial^{2} L(\beta, \phi)}{\partial \phi \partial \phi^{\prime}} = (w^{\prime} \diag(\mu(x \beta + w \phi) \odot (1 - \mu(x \beta + w \phi))) w + c_{1} v v^{\prime}) / \sqrt{NT} \, .
\end{equation}
Finally, note that specific choices for $c_{1}$ and the scaling factor are important for our asymptotic analysis.

To understand the rank deficiency problem problem and the derivation of the constraints, it is instructive to have a closer look at the linear index, $x_{ijt}^{\prime} \beta + \alpha_{it} + \gamma_{jt} + \rho_{ij}$. The incidental parameters enter additively into the linear index which makes the log-likelihood invariant to certain parameter transformations. For example, the linear index is invariant to adding a constant $c_{t}$ to all $\alpha_{it}$ while subtracting it from all $\gamma_{jt}$. Therefore, we introduce $T$ constraints $\sum_{i = 1}^{I} \alpha_{it} = \sum_{j = 1}^{J} \gamma_{jt}$ for $t = \{1, \ldots, T\}$, or in matrix notation $(1_{I} \otimes \eye_{T})^{\prime} \alpha = (1_{J} \otimes \eye_{T})^{\prime} \gamma$. Similarly, subtracting a constant $c_{i}$ from all $\alpha_{it}$ while adding it to all $\rho_{ij}$, or adding a constant $c_{j}$ to all $\gamma_{jt}$ while subtracting it from all $\rho_{ij}$ leaves the linear index unaffected, leading to $I$ constraints $\sum_{t = 1}^{T} \alpha_{it} = \sum_{j = 1}^{J} \rho_{ij}$ for $i = \{1, \ldots, I\}$ and $J$ constraints $\sum_{t = 1}^{T} \gamma_{it} = \sum_{i = 1}^{I} \rho_{ij}$ for $j = \{1, \ldots, J\}$. In matrix notation, these constraints translate to $(\eye_{I} \otimes 1_{T})^{\prime} \alpha = (\eye_{I} \otimes 1_{J})^{\prime} \rho$ and $(\eye_{J} \otimes 1_{T})^{\prime} \gamma = (1_{I} \otimes \eye_{J})^{\prime} \rho$. To impose all constraints simultaneously, we define 
\begin{equation}
    \label{eq:constraints}
	v = \begin{pmatrix}
        1_{I} \otimes \eye_{T}   & \eye_{I} \otimes 1_{T}     & 0_{IT \times J}	\\
        - 1_{J} \otimes \eye_{T} & 0_{JT \times I}         & \eye_{J} \otimes 1_{T} \\
        0_{IJ \times T}      & - \eye_{I} \otimes 1_{J}   & - 1_{I} \otimes \eye_{J}
	\end{pmatrix}
\end{equation}
such that $v^{\prime} \phi = 0$ characterizes the system of linear equality constraints. Because one of the constraints in $v$ is implied by all other constraints, the rank of $v$ reduces to $T + I + J - 1$. However, for our asymptotic analysis, it is more convenient to work with the $(IT + JT + IJ) \times (T + I + J)$ matrix $v$. In practice there can be several choices for $v$ that work. Perhaps the most familiar way is to set specific incidental parameters to zero, like excluding one time effect in models with individual and time effects without common intercept for classical panels. As in \textcite{fw2016}, for our asymptotic analysis it is however important to choose a specific normalization which is easier to work with.\footnote{More precisely, our normalization ensures that the inverse of the incidental parameter Hessian, defined in \eqref{eq:ip_hessian}, becomes block diagonal which helps us to bound its spectral norm in Lemma \ref{lemma:ip_hessian}.}

Since our primary interest is the estimation of the model parameters $\beta$, i.e.\ we treat the incidental parameters $\phi$ as high-dimensional nuisance parameters, we define the (profile) maximum likelihood estimator as 
\begin{equation}
    \label{eq:uncorrected_estimator}
    \hat{\beta} = \underset{\{\beta \in \mathbb{R}^{K}\}}{\argmin} \, L(\beta, \hat{\phi}(\beta)) \, , \quad
    \hat{\phi}(\beta) = \underset{\{\phi \in \mathbb{R}^{IT + JT + IJ}\}}{\argmin} \, L(\beta, \phi) \, .
\end{equation}

\begin{remark}[Computation]
    In empirical applications, since $K + IT + JT + IJ$ parameters have to be estimated jointly, \eqref{eq:uncorrected_estimator} quickly becomes a high-dimensional optimization problem. Consequently, using standard software routines that simply rely on generating $w$ for estimation is impractical, if not infeasible, even for moderately large panels. Therefore, we suggest the use of algorithms such as \textcite{gp2010}, \textcite{b2018}, \textcite{s2018}, or \textcite{cgz2019}, which are specifically designed to deal with this type of high-dimensional optimization problem. An example of ready-to-use software for fixed effects logit models, such as those analyzed in this paper, is the R package alpaca, which is based on the algorithm proposed in \textcite{s2018} and also provides the bias correction derived in this paper.
\end{remark}

\section{Asymptotic theory}
\label{sec:asymptotic_theory}

In this section, we derive the asymptotic properties of the maximum likelihood estimator $\hat{\beta}$, defined in \eqref{eq:uncorrected_estimator}, using an asymptotic framework where all three panel dimensions, $I$, $J$, and $T$, simultaneously grow to infinity. To simplify the notation and make our asymptotic analysis more concise, we follow \textcite{wz2021} and set $N = I = J$.

\subsection{Assumptions}

We make the following assumptions.
\begin{assumption}[Sampling and regularity conditions for three-dimensional panel binary logit models]~ %%% <-  Note that space! (Fix to force linebreak)
    \begin{enumerate}[i)]
    \item Sampling: The binary response $y_{ijt}$ is independently distributed over $i, j, t, N, T$ conditional on $\mathcal{F} \coloneqq \{x_{ijt}, \alpha_{it}, \gamma_{jt}, \rho_{ij} \, \colon \, i, j \in \{1, \ldots, N\}, \, t \in \{1, \ldots, T\}\}$.
    \item Model: For all $i, j \in \{1, \ldots, N\}$ and $t \in \{1, \ldots, T\}$,
    \begin{equation*}
        y_{ijt} = \ind\{x_{ijt}^{\prime} \beta + \alpha_{it} + \gamma_{jt}+ \rho_{ij} \geq \epsilon_{ijt}\} \, , \quad \epsilon_{ijt} \mid \mathcal{F} \sim F_{\epsilon} \, ,
    \end{equation*}
    where $F_{\epsilon}$ is the logistic cumulative distribution function. The realizations of the parameters and unobserved effects that generate the observed data are denoted by $\beta^{0}$ and $\phi^{0} = (\alpha^{0}, \gamma^{0}, \rho^{0})$. The unobserved effects $\phi^{0}$ are normalized to $v^{\prime} \phi^{0} = 0$.
    \item Compactness: The support of $x$, $\alpha^{0}$, $\gamma^{0}$, and $\rho^{0}$ is uniformly bounded over $i, j, t, N, T$.
    \item Non-collinearity: The explanatory variables $x_{ijt}$ are non-collinear after projecting out the unobserved effects, i.e.\
    \begin{equation*}
        \underset{\{\Delta \in \mathbb{R}^{K} \colon \norm{\Delta} = 1\}}{\min} \; \underset{\{\pi \in \mathbb{R}^{2NT + N^2}\}}{\min} \;  \frac{1}{N^2T} \sum_{i = 1}^{N} \sum_{j = 1}^{N} \sum_{t = 1}^{T} (x_{ijt} \Delta - w_{ijt} \pi)^{2} \geq c_{2} \, ,
    \end{equation*}
    where $0 < c_{2} < \infty$ is a finite constant independent of the sample size. 
    \item Asymptotics: We consider limits of sequences where $N / T \rightarrow c_{3}$ with $0 < c_{3} < \infty$ as $N, T \rightarrow \infty$.
    \end{enumerate}
\end{assumption}

\begin{remark}[Assumption 1]
    i) restricts the distribution of the outcome variable, conditional on the explanatory variables and the unobserved effects. A similar assumption has been used by \textcite{hn2004}, for classical panels, and it is a natural starting point for our asymptotic analysis. Moreover, our asymptotic analysis also holds for panels where $I$ and $J$ are of different sizes, as long as $I \sim J = \mathcal{O}(N)$. ii) requires the explanatory variables to be strictly exogenous. This assumption rules out any form of feedback from past realizations of the binary outcome variables to the explanatory variables, e.g.\ it rules out functions of lagged outcome variables as regressors. Further, we restrict our analysis to logit models for analytical convenience. We expect that our results can be generalized to weakly exogenous explanatory variables, e.g.\ lagged outcome variables, and other cumulative distribution functions, e.g.\ if $F_{\epsilon}$ is the standard normal cumulative distribution function as assumed in probit models. However, this comes at the cost of more involved proofs along with different assumptions about the dependence over time. Our conjecture is further supported by \textcite{hsw2020}, who studied bias corrections for a dynamic probit version of our model via simulation experiments. Moreover, our model implies certain Bartlett identities that can be used to simplify the bias expressions, as suggested by \textcite{f2009}. iii) is a compact support assumption, as in \textcite{fw2016}, and ensures that $0 < \mu_{ijt}(\beta^{0}, \phi^{0}) < 1$ for all $i, j, t, N, T$. Thus, in the terminology of the network literature, we implicitly assume that the network of binary decisions is sufficiently dense over time. iv) imposes restrictions on the explanatory variables used in the model. That is, only regressors that vary across all three panel dimensions may be included in the model. v) establishes the asymptotic framework used in our analysis. As in \textcite{fw2016}, all panel dimensions have to grow large at a constant relative rate to derive a non-degenerate asymptotic distribution for the debiased estimator.
\end{remark}

\begin{remark}[Missing observations]
    In empirical applications it is quite common that some observations are missing due to some attrition process. However, as noted by \textcite{fw2018}, this does not affect the asymptotic analysis, apart from introducing inconvenience due to additional notation, as long as the attrition process is random, conditional on $\mathcal{F}$, and there is only a fixed number of missing observations for each $i$, $j$, and $t$. For example, in three-dimensional panels used in international trade (see Example \ref{ex:bilateral} in Section \ref{sec:model_and_estimator}), usually observations where $i = j$ are missing, as countries do not trade with themselves. The conditions of \textcite{fw2018} hold in this example because the attrition process is deterministic and there is only one missing observation for each $i$ and $j$.
\end{remark}

\subsection{Asymptotic distribution}

Before presenting the asymptotic distribution, we first need to introduce some additional notation. Let $\mu^{\langle 1 \rangle}(z) = \partial_{z} \mu(z)$, $\mu^{\langle 2 \rangle}(z) = \partial_{z^2} \mu(z)$, and $\mu^{\langle 3 \rangle}(z) = \partial_{z^3} \mu(z)$ denote the first-, second-, and third-order derivatives of the logistic cumulative distribution function $\mu(\cdot)$. Further, we define $\mu_{ijt} = \mu(x_{ijt}^{\prime} \beta^{0} + w_{ijt}^{\prime} \phi^{0})$. The definitions of $\mu_{ijt}^{\langle 1 \rangle}$, $\mu_{ijt}^{\langle 2 \rangle}$, and $\mu_{ijt}^{\langle 3 \rangle}$ follow accordingly. For every regressor $x_{k}$, we define $\tilde{x}_{k} = x_{k} - w^{\prime} \phi_{k}^{\ast}$, where
\begin{equation*}
    \phi_{k}^{\ast} = \underset{\{\phi_{k} \in \mathbb{R}^{2NT + N^2}\}}{\argmin} \;  \frac{1}{\sqrt{NT}} \sum_{i = 1}^{N} \sum_{j = 1}^{N} \sum_{t = 1}^{T} \mu_{ijt}^{\langle 1 \rangle} (x_{ijt, k} - w_{ijt} \phi_{k})^{2} 
\end{equation*}
are the coefficients of a weighted least-squares problem. The residuals $\tilde{x}_{k}$ stem from Legendre transforms that we use to project out the incidental parameters from the asymptotic expansions (details about the transformation are provided in Appendix \ref{app:asymptotic_expansions}). Furthermore, we define the leading asymptotic bias 
\begin{equation}
    \label{eq:leading_bias}
    b = \overline{W}^{- 1} (B_{\alpha} + B_{\gamma} + B_{\rho}) \, ,
\end{equation}
where
\begin{equation}
    \label{eq:profile_hessian}
    W = \frac{1}{N^2 T} \sum_{i = 1}^{N} \sum_{j = 1}^{N} \sum_{t = 1}^{T} \mu_{ijt}^{\langle 1 \rangle} \, \tilde{x}_{ijt} \, \tilde{x}_{ijt}^{\prime}
\end{equation}
is the normalized profile Hessian, $\overline{W} = \EX{W}$, and
\begin{align*}
    B_{\alpha} =& \, - \frac{1}{2 NT} \sum_{i = 1}^{N} \sum_{t = 1}^{T} \frac{\sum_{j = 1}^{N} \mu_{ijt}^{\langle 2 \rangle} \tilde{x}_{ijt}}{\sum_{j = 1}^{N} \mu_{ijt}^{\langle 1 \rangle}} \, , \\
    B_{\gamma} =& \, - \frac{1}{2 NT} \sum_{j = 1}^{N} \sum_{t = 1}^{T} \frac{\sum_{i = 1}^{N} \mu_{ijt}^{\langle 2 \rangle} \tilde{x}_{ijt}}{\sum_{i = 1}^{N} \mu_{ijt}^{\langle 1 \rangle}} \, , \\
    B_{\rho} =& \, - \frac{1}{2 N^2} \sum_{i = 1}^{N} \sum_{j = 1}^{N} \frac{\sum_{t = 1}^{T} \mu_{ijt}^{\langle 2 \rangle} \tilde{x}_{ijt}}{\sum_{t = 1}^{T} \mu_{ijt}^{\langle 1 \rangle}} \, ,
\end{align*}
are bias components that arise from estimating the incidental parameters $\alpha$, $\gamma$, and $\rho$, respectively.

We establish in the following Theorem that $\hat{\beta}$ has a degenerating asymptotic distribution.
\begin{theorem}[Degenerating asymptotic distribution of the uncorrected estimator]\label{theorem:uncorrected}
    Let Assumptions 1 and 2 hold. Then,
    \begin{equation*}
        N\sqrt{T} (\hat{\beta} - \beta^{0}) = W^{- 1} U^{(0)} + \mathcal{O}_{P}(\max(\sqrt{N}, \sqrt{T})) \, ,
    \end{equation*}
    where
    \begin{equation*}
         W^{- 1} U^{(0)} \overset{d}{\rightarrow} \mathcal{N}(0, \overline{W}^{- 1})
    \end{equation*}
    with
    \begin{equation*}
        U^{(0)} =  \frac{1}{N \sqrt{T}} \sum_{i = 1}^{N} \sum_{j = 1}^{N} \sum_{t = 1}^{T} \tilde{x}_{ijt} (y_{ijt} - \mu_{ijt}) \, ,
    \end{equation*}
    $W$ is the normalized profile Hessian defined in \eqref{eq:profile_hessian}, and $\overline{W} = \EX{W} > 0$.
\end{theorem}

We proof Theorem \ref{theorem:uncorrected} in Appendix \ref{app:theorem_uncorrected}.

\begin{remark}[Theorem \ref{theorem:uncorrected}]
    Contrary to the results from previous literature, e.g.\ \textcite{hn2004}, \textcite{fw2016}, or \textcite{wz2021}, the uncorrected estimator has a degenerating asymptotic distribution. For the asymptotic distribution to have a constant bias, both sides would have to be divided by $\max(\sqrt{N}, \sqrt{T})$. However, this would cause the asymptotic covariance matrix $\overline{W}^{- 1}$ to shrink towards zero. Thus, the order of the bias and variance cannot be balanced to obtain a non-degenerate asymptotic distribution. This result is the consequence of a more severe imbalance between the convergence rates of $\hat{\beta}$ and $\hat{\phi}(\hat{\beta})$ than reported in the previous literature. The convergence rate of $\hat{\beta}$ is $N \sqrt{T}$, while the convergence rate of $\hat{\phi}(\hat{\beta})$ is $\max(\sqrt{N}, \sqrt{T})$.
\end{remark}

The following theorem states that after correcting the leading asymptotic bias, we obtain a correctly centered non-degenerate asymptotic distribution.
\begin{theorem}[Asymptotic distribution of the bias-corrected estimator]\label{theorem:debiased}
    Let Assumption 1 hold. Then,
    \begin{equation*}
        N\sqrt{T} (\hat{\beta} - \beta^{0} - b / \sqrt{NT}) \overset{d}{\rightarrow} \mathcal{N}(0, \overline{W}^{- 1}) \, .
    \end{equation*}
\end{theorem}

The proof of Theorem \ref{theorem:debiased} is provided in Appendix \ref{app:theorem_debiased}.

\begin{remark}[Theorem \ref{theorem:debiased}]
    Because the normalized leading asymptotic bias $b / \sqrt{NT}$ is of the form $B_{\alpha} / N + B_{\gamma} / N + B_{\rho} / T$, the order of the bias $\max(N^{-1}, T^{-1})$ is always larger than the order of the standard deviation $(N\sqrt{T})^{- 1}$. Further, our results support the conjecture of \textcite{fw2018}, which is based on a heuristic formula. Their heuristic correctly predicts that our uncorrected estimator has a bias of order,
    \begin{equation*}
       \dim(\phi) / (N^2T) = (2NT + N^2)/(N^2T) = 1 / N + 1 / N + 1 / T \, .
    \end{equation*}
\end{remark}

\subsection{Bias correction}

Theorem \ref{theorem:uncorrected} shows that the uncorrected estimator has a degenerating asymptotic distribution. Consequently, standard maximum likelihood inference, i.e.\ confidence regions constructed around the uncorrected estimator are in general invalid. However, the inference problem can be resolved, as shown in Theorem \ref{theorem:debiased}, by subtracting the leading asymptotic bias $b$ defined in \eqref{eq:leading_bias}.

In the following, we use plug-in estimates of $B_{\alpha}$, $B_{\gamma}$, $B_{\rho}$, and $\overline{W}$ to construct a bias-corrected estimator $\tilde{\beta}$. Let $\hat{\mu}_{ijt} = \mu(x_{ijt}^{\prime} \hat{\beta} + w_{ijt}^{\prime} \hat{\phi})$, $\hat{\mu}_{ijt}^{\langle 1 \rangle} = \mu^{\langle 1 \rangle}(x_{ijt}^{\prime} \hat{\beta} + w_{ijt}^{\prime} \hat{\phi})$, and $\hat{\mu}_{ijt}^{\langle 2 \rangle} = \mu^{\langle 2 \rangle}(x_{ijt}^{\prime} \hat{\beta} + w_{ijt}^{\prime} \hat{\phi})$. Further, for every regressor $x_{k}$, we define $\hat{\tilde{x}}_{k} = x_{k} - w \hat{\phi}_{k}^{\ast}$, where
\begin{equation}
    \label{eq:x_tilde_hat}
    \hat{\phi}_{k}^{\ast} = \underset{\{\phi_{k} \in \mathbb{R}^{2NT + N^2}\}}{\argmin} \; \frac{1}{\sqrt{NT}} \sum_{i = 1}^{N} \sum_{j = 1}^{N} \sum_{t = 1}^{T} \hat{\mu}_{ijt}^{\langle 1 \rangle} (x_{ijt, k} - w_{ijt} \phi_{k})^{2} \, .
\end{equation}
are the coefficients of a weighted least-squares problem. Then, a bias-corrected estimator is constructed as
\begin{equation}
    \label{eq:debiased_estimator}
    \tilde{\beta} = \hat{\beta} - \widehat{W}^{- 1} (\widehat{B}_{\alpha} / N + \widehat{B}_{\gamma} / N + \widehat{B}_{\rho} / T) \, ,
\end{equation}
where
\begin{align}
        \widehat{B}_{\alpha} =& \, - \frac{1}{2 NT} \sum_{i = 1}^{N} \sum_{t = 1}^{T} \frac{\sum_{j = 1}^{N} \hat{\mu}_{ijt}^{\langle 2 \rangle} \, \hat{\tilde{x}}_{ijt}}{\sum_{j = 1}^{N} \hat{\mu}_{ijt}^{\langle 1 \rangle}} \, , \label{eq:estimators_bias_variance} \\
        \widehat{B}_{\gamma} =& \, - \frac{1}{2 NT} \sum_{j = 1}^{N} \sum_{t = 1}^{T} \frac{\sum_{i = 1}^{N} \hat{\mu}_{ijt}^{\langle 2 \rangle} \, \hat{\tilde{x}}_{ijt}}{\sum_{i = 1}^{N} \hat{\mu}_{ijt}^{\langle 1 \rangle}} \, , \nonumber \\
        \widehat{B}_{\rho} =& \, - \frac{1}{2 N^2} \sum_{i = 1}^{N} \sum_{j = 1}^{N} \frac{\sum_{t = 1}^{T} \hat{\mu}_{ijt}^{\langle 2 \rangle} \, \hat{\tilde{x}}_{ijt}}{\sum_{t = 1}^{T} \hat{\mu}_{ijt}^{\langle 1 \rangle}} \, , \nonumber \\
        \widehat{W} =& \, \frac{1}{N^2 T} \sum_{i = 1}^{N} \sum_{j = 1}^{N} \sum_{t = 1}^{T} \hat{\mu}_{ijt}^{\langle 1 \rangle} \, \hat{\tilde{x}}_{ijt} \, \hat{\tilde{x}}_{ijt}^{\prime} \, , \nonumber
\end{align}
and $\hat{\beta}$ is the uncorrected maximum likelihood estimator defined in \eqref{eq:uncorrected_estimator}.

The following Lemma shows that the estimators for the various bias components and the expected normalized profile Hessian, proposed in \eqref{eq:estimators_bias_variance}, are consistent.
\begin{lemma}[Consistency of estimators for bias and variance components]\label{lemma:consistency_bias_variance}
     Let Assumption 1 hold. Then,
     \begin{equation*}
        \norm{\widehat{B}_{\alpha} - B_{\alpha}}_{2} = o_{P}(1) \, ,
        \norm{\widehat{B}_{\gamma} - B_{\gamma}}_{2} = o_{P}(1) \, ,
        \norm{\widehat{B}_{\rho} - B_{\rho}}_{2} = o_{P}(1) \, , 
        \norm{\widehat{W} - \overline{W}}_{2} = o_{P}(1) \, .
     \end{equation*}
\end{lemma}

We proof Lemma \ref{lemma:consistency_bias_variance} in Appendix \ref{app:lemma_consistency_bias_variance}.

\begin{remark}[Uninformative observations]
    A particular problem that arises in empirical applications of nonlinear fixed effects models are uninformative observations. In binary choice models, observations become uninformative whenever a subset of the outcome variable needed to estimate one of the incidental parameters is either a vector of zeros or ones, i.e.\ a vector without variation. For example, if a pair $(ij)$ never changes the status of the dependent variable over the entire time horizon, the corresponding estimate for $\rho_{ij}$ does not exist and thus the respective observations cannot contribute to the estimation of $\beta$ or to any of the other incidental parameters. Consequently, these observations are generally uninformative and can be removed without affecting the estimation results. Importantly, in our setting, removing uninformative observations can cause the data set to become unbalanced. \textcite{cs2022} denote this phenomenon as latent unbalancedness and analyze its implications for the finite sample performance of these estimators. Intuitively, the incidental parameter estimates are more sensitive to the removal of observations than the estimates of $\beta$. This is because the incidental parameter estimates are based on a smaller number of observations. Consequently, the inference problem is further amplified. Finally, we would like to point out that for numerical reasons, uninformative observations should be removed from the sample. Keeping these observations can slow down the convergence of the optimization routine or even cause it to fail. Moreover, the corresponding estimates of the incidental parameters will be very large in absolute value and dominate the linear index of the corresponding uninformative observations. Because the linear index enters numerators and denominators when estimating the bias components, these inflated linear indices can cause numerical problems and contaminate the estimates of the bias components.
\end{remark}

\begin{remark}[Jackknife and bootstrap bias corrections]
    Although we do not explicitly analyze other bias corrections, we expect that jackknife and bootstrap bias corrections can be applied as well given the derived form of the normalized bias, $B_{\alpha} / N + B_{\gamma} / N + B_{\rho} / T$. For example, the form of the bias suggests that a split-panel jackknife bias-corrected estimator, akin to \textcite{dj2015} and \textcite{fw2016}, can be constructed by forming suitable half-panels along each of the three panel dimensions. \textcite{hsw2020} consider such an approach in their numerical exercise and provide explicit formulas. However, it is important to note that split-panel jackknife bias corrections require an additional conditional homogeneity assumption similar to Assumption 4.3 in \textcite{fw2016}. See \textcite{fw2018} for other jackknife and bootstrap bias corrections, like, among others, the leave-one out jackknife of \textcite{hn2004} or the $k$-step bootstrap of \textcite{ks2016}.
\end{remark}

\begin{remark}[Computation continued]
    Like the estimation of $\beta$, bias corrections become computationally demanding when $N$ and/or $T$ become large. For the jackknife and bootstrap bias corrections, the computational burden arises from the need to re-estimate $\beta$ for different (sub)samples of the original data set. The computational challenge for the bias correction proposed in this paper are the residuals $\hat{\tilde{x}}$ of a high-dimensional optimization problem defined in \eqref{eq:x_tilde_hat}. \textcite{cs2019} explain the efficient computation of $\hat{\tilde{x}}$ using the example of the analytical bias correction of \textcite{fw2016}. We provide a computationally efficient version of our bias correction in the R package alpaca.
\end{remark}

\section{Differences to previous results in the literature}
\label{sec:differences}

To better align the results presented in the previous section with the results from the previous literature, we compare our results with the ones from the two most related papers, \textcite{fw2016} and \textcite{wz2021}, and discuss where the differences come from.

\subsection{General challenges with multi-way fixed effects}

Compared to classical panel models with only individual fixed effects, e.g.\ \textcite{hn2004}, \textcite{f2009}, and \textcite{hk2011}, models with additional fixed effects add further complications to the asymptotic analysis. First, it is not possible to express the log-likelihood function as a sum of individual log-likelihood contributions, where each log-likelihood contribution depends only on a fixed-dimensional set of parameters. This strategy was proposed by \textcite{hn2004} to deal with the infinite-dimensional parameter space and is a common strategy in the panel data literature. Second, the incidental parameter Hessian $\partial_{\phi \phi^{\prime}} L(\beta, \phi)$ is no longer diagonal, which complicates, for example, bounding some quantities in the asymptotic expansion. \textcite{fw2016} solve both issues for classical panel data models with individual and time effects. For the first issue, they propose a projection method based on Legendre transforms of the log-likelihood function to obtain asymptotic expansions which do not depend on the incidental parameters. For the second issue, they establish an approximation argument for the inverse of the incidental parameter Hessian, in which they show that asymptotically the inverse is a (weakly) diagonally dominant matrix (see Lemma D.1 in \textcite{fw2016}). Thus, it can be uniformly approximated by a diagonal matrix and has off-diagonal elements that are sufficiently small. This approximation argument is particularly important to show that the asymptotic bias can be ``decoupled'', i.e.\ expressed as the sum of two bias components, one for each set of fixed effects in the model specification. Finally, the most related paper is \textcite{wz2021}, that analyzes the properties of a fixed effects (Pseudo-)Poisson estimator for three-dimensional panels with the same linear index specification as in this paper under similar assumptions. However, contrary to us, the authors can exploit a unique property of the (Pseudo-)Poisson model that allows them to profile-out $\rho$ from the (pseudo-)log-likelihood function. This essentially turns their three-way model into a stacked two-way model, with only $\alpha$ and $\gamma$ as incidental parameters, and allows them to rely on the results of \textcite{fw2016} for their asymptotic analysis. As a consequence, \textcite{wz2021} only need $N = I = J$ to grow to infinity and $T$ can be fixed.

\subsection{Differences in asymptotic distributions}

Both, \textcite{fw2016} and \textcite{wz2021}, derive non-degenerate asymptotic distributions of $\hat{\beta}$,
\begin{equation*}
    r_{n} (\hat{\beta} - \beta^{0} - b / r_{n}) \overset{d}{\rightarrow} \mathcal{N}(0, V) \, ,
\end{equation*}
where $r_{n}$ is the convergence rate of $\hat{\beta}$, $b$ is the constant leading bias, and $V$ is an asymptotic covariance matrix. Here, the order of the normalized asymptotic bias $b / r_{n} = \mathcal{O}_{P}(r_{n}^{- 1})$ and the convergence rate of $\hat{\beta}$ are exactly balanced yielding a non-degenerate but distorted asymptotic distribution.
\begin{figure}[!htbp]
    \caption{Asymptotic distribution of $\hat{\beta}$ derived in \textcite{fw2016} and \textcite{wz2021}}
    \centering
	\includegraphics[width=0.9\textwidth]{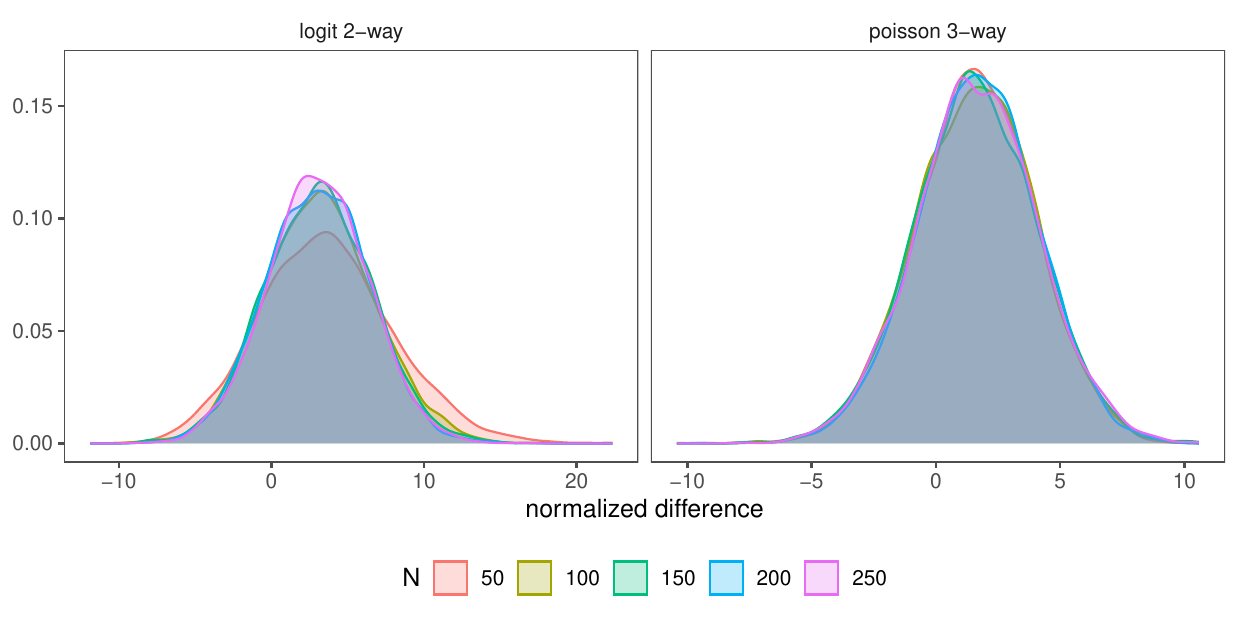}
	\begin{minipage}{0.9\textwidth}
	\footnotesize
	\textbf{Notes:} The left panel is based on the data generating process of \textcite{fw2016} for static panel probit models adapted to logit models, with $T(N) = N / 5$; the right panel is based on DGP I of \textcite{wz2021}, with $N = I = J$ and $T = 5$; in the left pane $r_{n} = \sqrt{NT}$ and in the right panel $r_{n} = N$ is used to normalize the differences; results are based on $5{,}000$ simulated samples for each $N$.\\
    \textbf{DGP (left panel):} The data generating process is $y_{it} = \ind \{\beta x_{it} + \alpha_{i} + \gamma_{j} \geq \log(u_{it} /(1 - u_{it}))\}$, $x_{it} = 0.5 x_{it-1} + \alpha_{i} + \gamma_{j} + v_{it}$, where $i = \{1,\ldots, N\}$, $t = \{1,\ldots, T\}$, $\alpha_{i}, \gamma_{j} \sim \iid \N(0, 1 / 16)$, $u_{it} \sim \iid \U(0, 1)$, $v_{it} \sim \iid \N(0, 1 / 2)$, $x_{i0} \sim \iid \N(0, 1)$, and $\beta = 1$.\\
    \textbf{DGP (right panel):} The data generating process is $y_{ijt} = \lambda_{ijt} \omega_{ijt}$, $\lambda_{ijt} = \exp(\beta x_{ijt} + \alpha_{it} + \gamma_{jt} + \rho_{ij})$, $\omega_{ijt} = \exp(- 0.5 \log(1 + \lambda_{ijt}^{- 2}) + (\log(1 + \lambda_{ijt}^{- 2}))^{1 / 2} z_{ijt}))$, $x_{ijt} = 0.5 x_{ijt-1} + \alpha_{it} + \gamma_{jt} + \rho_{ij} + v_{ijt}$, $z_{ijt} = 0.3 z_{ijt-1} + u_{ijt}$, where $i, j = \{1,\ldots, N\}$, $t = \{1,\ldots, T\}$, $\alpha_{it}, \gamma_{jt}, \rho_{ij} \sim \iid \N(0, 1 / 256)$, $v_{ijt} \sim \iid \N(0, 1 / 4)$, $u_{ijt} \sim \iid \N(0, 0.91)$, $x_{ij0} = \rho_{ij} + v_{ij0}$, $z_{ij0} \sim \iid \N(0, 1)$, and $\beta = 1$.
    \end{minipage}
    \label{fig:simulated_distribution_others}
\end{figure}
Figure \ref{fig:simulated_distribution_others} illustrates the non-degenerate asymptotic distributions derived in both papers. It shows the empirical densities of the normalized differences of the uncorrected estimators and the true parameter values, $r_{n} (\hat{\beta} - \beta^{0})$, for a logit model with individual and time effects (left panel), as studied in \textcite{fw2016}, and for a Pseudo-Poisson model with exporter-time, importer-time, and exporter-importer effects (right panel), as studied in \textcite{wz2021}. The figure is based on simulated data for different sample sizes. We use the data generating processes from the corresponding papers. The figure shows that as the sample size increases, the distribution of the uncorrected estimator converges to a normal distribution centered around the bias. Both papers propose bias corrections to re-center the asymptotic distribution properly to ensure reliable inference.

In contrast, Theorem 2 reveals that in our case, the order of the normalized asymptotic bias $b / \sqrt{NT} = \mathcal{O}_{P}((NT)^{- 1 / 2})$ and the convergence rate of $\hat{\beta}$, $r_{n} = N\sqrt{T}$, are not balanced. More precisely, the normalized bias shrinks slower than the standard deviation of the uncorrected estimator, resulting in a degenerating asymptotic distribution.
\begin{figure}[!htbp]
    \caption{Asymptotic distribution of $\hat{\beta}$ derived in this paper under different normalization}
    \centering
	\includegraphics[width=0.9\textwidth]{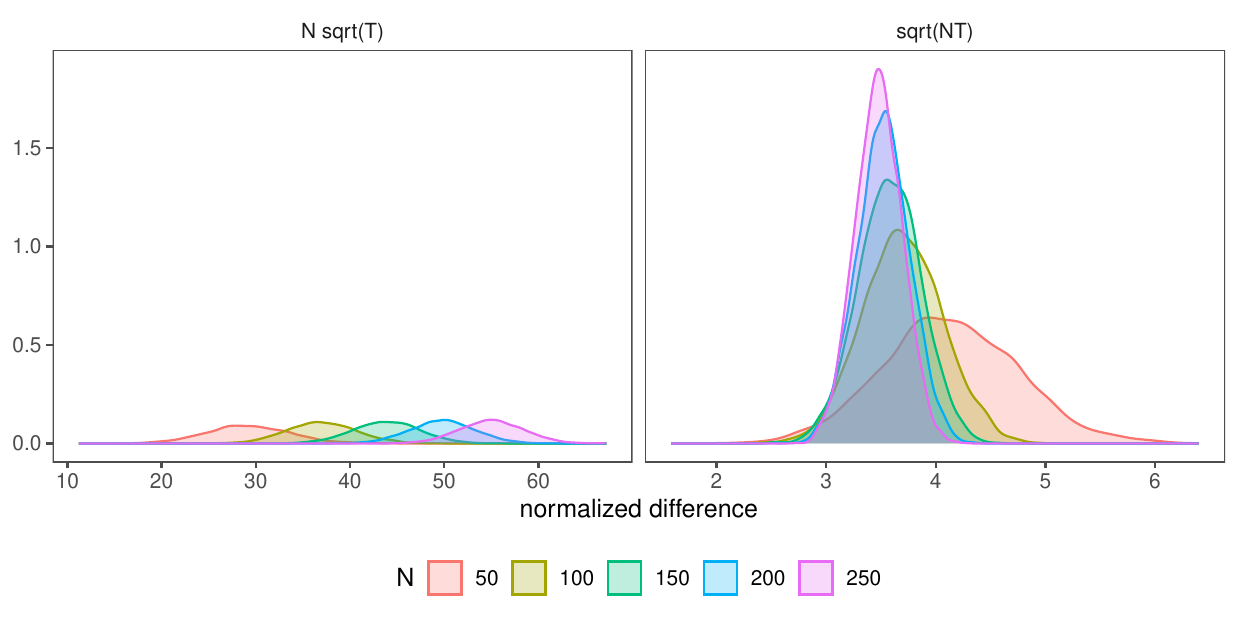}
	\begin{minipage}{0.9\textwidth}
	\footnotesize
	\textbf{Notes:} Both panels are based on the data generating process described in equation \eqref{eq:simulation} in Section \ref{sec:simulation_experiments}, with $T(N) = N / 5$; in the left panel the normalizing constant is $N \sqrt{T}$ and in the right panel the normalizing constant is $\sqrt{NT}$; results are based on $5{,}000$ simulated samples for each $N$.
    \end{minipage}
    \label{fig:logit_3way}
\end{figure}
Figure \ref{fig:logit_3way} illustrates the balancing problem using two different normalizing constants for the differences of the uncorrected estimators and the true parameter values: (left panel) $N \sqrt{T} (\hat{\beta} - \beta^{0})$ and (right panel) $\sqrt{NT} (\hat{\beta} - \beta^{0})$. The figure shows empirical densities of the normalized differences based on simulated data for different sample sizes. The data generation process is introduced in equation \eqref{eq:simulation} in Section \ref{sec:simulation_experiments}. The left panel shows that the mean of the normalized differences increases with the sample size, while the variance converges to a constant. The right panel shows the opposite: the mean converges to a constant, but the variance decreases to zero. This illustrates that, unlike Figure \ref{fig:simulated_distribution_others}, there is no appropriate normalization that yields a non-degenerate asymptotic distribution. Therefore, unlike \textcite{fw2016} and \textcite{wz2021}, the order of bias and variance cannot be exactly balanced. However, as shown in Theorem \ref{theorem:debiased}, we can construct an estimator with a correctly centered non-degenerate asymptotic distribution.

\subsection{Differences in incidental parameter Hessians}

\begin{figure}[!htbp]
    \caption{Incidental parameter Hessian -- two- and three-way fixed effects}
    \centering
	\includegraphics[width=0.9\textwidth]{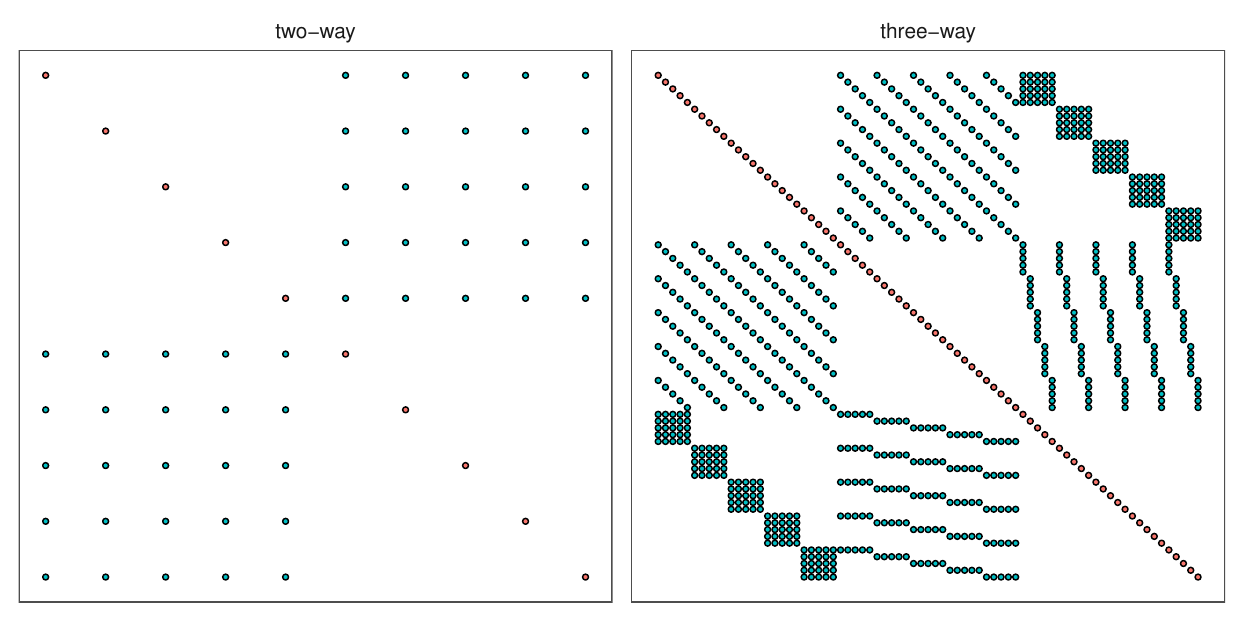}
	\begin{minipage}{0.9\textwidth}
	\footnotesize
	\textbf{Notes:} Dots indicate positive entries in both matrices; red and green dots denote higher and lower order values, respectively; $N = T = 5$. The linear index in the two-way model is $x_{it}^{\prime} \beta + \alpha_{i} + \gamma_{t}$ and the vector collecting the incidental parameters is $\phi = (\alpha, \gamma)$. The linear index in our three-way model is $x_{ijt}^{\prime} \beta + \alpha_{it} + \gamma_{jt} + \rho_{ij}$ and the vector collecting the incidental parameters is $\phi = (\alpha, \gamma, \rho)$. In $\alpha$ and $\gamma$ the ordering of the single parameters is such that, time runs faster than $i$ or $j$ respectively, and in $\rho$ the $j$-index runs faster than the $i$-index.
    \end{minipage}
    \label{fig:ip_hessian_sparsity}
\end{figure}

Properly handling the incidental parameter Hessian, which enters asymptotic expansions through its inverse, is key to the strategy of \textcite{fw2016} and therefore also to \textcite{wz2021}. It is important to bound certain quantities in asymptotic expansions and to ensure that the asymptotic bias can be decoupled into separate bias components. Figure \ref{fig:ip_hessian_sparsity} shows the structure of the incidental parameter Hessians without constraints, $\partial_{\phi \phi^{\prime}} L_{u}(\beta, \phi)$, from \textcite{fw2016} (left panel) and from this paper (right panel), for a data set with $N = T = 5$. For ease of exposition, we look at the unconstrained incidental parameter Hessians, as the difference is already apparent here.
  We do not additionally show the Hessian of \textcite{wz2021}, as their proof strategy is fundamentally based on \textcite{fw2016}. The left Hessian is of dimension $(N + T) \times (N + T)$ and the right Hessian is of dimension $(2NT + N^2) \times (2NT + N^2)$. Higher order values (red dots) are located on the diagonals of the matrices, while all non-zero off-diagonal values are of lower order (green dots). Although the right Hessian has a much higher dimension than the left Hessian, both matrices have asymptotically the same number of non-zero off-diagonal elements. The main difference between our Hessian and the one derived by \textcite{fw2016} is that our Hessian has sparse off-diagonal blocks, while theirs has dense off-diagonal blocks. This sparsity pattern is due to the overlapping fixed effects in our model specification. For example, $\partial_{\alpha_{it} \alpha_{i^{\prime} t^{\prime}}} L_{u}(\beta, \phi) = \sum_{j = 1}^{N} \mu^{\langle 1 \rangle}_{ijt} / \sqrt{NT}$ if $i = i^{\prime}$ and  $t = t^{\prime}$ and zero otherwise, because  $\alpha_{it}$ and $\alpha_{i^{\prime} t^{\prime}}$ only enter in the same linear index if $i = i^{\prime}$ and $t = t^{\prime}$. This leads to the $NT \times NT$ diagonal block, $\partial_{\alpha \alpha^{\prime}} L_{u}(\beta, \phi)$, with elements of order $N / \sqrt{NT}$. The two other diagonal blocks, $\partial_{\gamma \gamma^{\prime}} L_{u}(\beta, \phi)$ and $\partial_{\rho \rho^{\prime}} L_{u}(\beta, \phi)$, follow analogously. Additionally, $\partial_{\alpha_{it} \gamma_{j t^{\prime}}} L_{u}(\beta, \phi) = \mu^{\langle 1 \rangle}_{ijt} / \sqrt{NT}$ if $t = t^{\prime}$ and zero otherwise, because $\alpha_{it}$ and $\gamma_{j t^{\prime}}$ only enter in the same linear index if $t = t^{\prime}$. This results in the sparse $NT \times NT$ off-diagonal block $\partial_{\alpha \gamma^{\prime}} L_{u}(\beta, \phi)$ with $NT$ elements of order $1 / \sqrt{NT}$. The other off-diagonal blocks, $\partial_{\alpha \rho^{\prime}} L_{u}(\beta, \phi)$, $\partial_{\gamma \alpha^{\prime}} L_{u}(\beta, \phi)$, $\partial_{\gamma \rho^{\prime}} L_{u}(\beta, \phi)$, $\partial_{\rho \alpha^{\prime}} L_{u}(\beta, \phi)$, and $\partial_{\rho \gamma^{\prime}} L_{u}(\beta, \phi)$, follow analogously. Intuitively, although a three-dimensional panel is much larger than a classical panel, the number of observations that can be used to estimate the incidental parameters is asymptotically the same as in a classical panel with individual and time effects. Thus, the increased sample size does not improve the convergence rates of the corresponding estimators, which are still $\sqrt{N}$ or $\sqrt{T}$. Importantly, the sparsity is also reflected in the inverse of the incidental parameter Hessian. Thus, properly handling this sparsity is one of the main challenges in deriving our results (see Appendixes \ref{app:ip_hessian} and \ref{app:linear_operator} for further details).

\section{Simulation experiments}
\label{sec:simulation_experiments}

In this section, we conduct simulation experiments to study the finite sample behaviour of the uncorrected and debiased maximum likelihood estimators of the model parameters defined in \eqref{eq:uncorrected_estimator} and \eqref{eq:debiased_estimator}, respectively. We analyze biases and the reliability of the derived asymptotic distributions for inference. In particular, we consider the following statistics for our analysis: relative bias in percent, bias relative to standard deviation, and coverage rates of confidence intervals with 95\% nominal level. We adapt the static data generating process of \textcite{fw2016} to logit models for bilateral panels with three sets of overlapping unobserved effects,
\begin{align}
\label{eq:simulation}
    y_{ijt} &= \, \ind \{\beta \, x_{ijt} + \alpha_{it} + \gamma_{jt} + \rho_{ij} \geq \log(u_{ijt} /(1 - u_{ijt}))\} \, ,\\
    x_{ijt} &= \, x_{ijt-1} \, / \, 2 + \alpha_{it} + \gamma_{jt} + \rho_{ij} + v_{ijt} \, , \nonumber
\end{align}
where $i, j = \{1,\ldots, N\}$, $t = \{1, \ldots, T\}$, $\alpha_{it}, \gamma_{jt}, \rho_{ij} \sim \iid \N(0, 1 / 24)$, $u_{ijt} \sim \iid \U(0, 1)$,  $v_{ijt} \sim \iid \N(0, 1 / 2)$, and $x_{ij0} \sim \iid \N(0, 1)$. We set $\beta = 1$ and generate data sets with $N \in \{50, 75, \dots, 225, 250 \}$ senders and receivers observed for $T(N) = N / 5 $ time periods. Our study design ensures that $N$ and $T$ grow at a constant rate and is therefore in line with our asymptotic analysis. All results presented are based on $5{,}000$ simulated samples for each $N$.

\begin{table}[!htbp]
\centering
\caption{Finite sample properties of estimators for model parameters}
\begin{threeparttable}
\begin{tabular}{lrrrrrr}
    \toprule
    $(N, T)$ & \multicolumn{3}{c}{uncorrected} & \multicolumn{3}{c}{debiased} \\
    \cmidrule(lr){2-4}\cmidrule(lr){5-7}    
    & Bias (in \%) & Bias / SD & Coverage & Bias (in \%) & Bias / SD & Coverage \\
    \midrule
    (50, 10) & 18.465 & 6.637 & 0.000 & -0.921 & -0.407 & 0.947 \\ 
    (75, 15) & 11.404 & 8.484 & 0.000 & -0.387 & -0.327 & 0.947 \\ 
    (100, 20) & 8.268 & 10.085 & 0.000 & -0.202 & -0.270 & 0.953 \\ 
    (125, 25) & 6.473 & 11.304 & 0.000 & -0.133 & -0.249 & 0.946 \\ 
    (150, 30) & 5.337 & 12.389 & 0.000 & -0.078 & -0.192 & 0.945 \\ 
    (175, 35) & 4.530 & 13.456 & 0.000 & -0.057 & -0.179 & 0.946 \\ 
    (200, 40) & 3.939 & 14.622 & 0.000 & -0.040 & -0.157 & 0.949 \\ 
    (225, 45) & 3.475 & 15.643 & 0.000 & -0.038 & -0.178 & 0.949 \\ 
    (250, 50) & 3.113 & 16.362 & 0.000 & -0.032 & -0.174 & 0.949 \\ 
    \bottomrule
\end{tabular}
\begin{tablenotes}
    \footnotesize
    \item\textbf{Notes:} uncorrected and debiased refer to estimates obtained from \eqref{eq:uncorrected_estimator} and \eqref{eq:debiased_estimator}, respectively; SD and Coverage indicate standard deviation and coverage rates of confidence intervals with 95\% nominal level, respectively; results based on $5{,}000$ simulated samples for each $(N, T)$.
\end{tablenotes}
\end{threeparttable}
\label{tab:simulation_results}
\end{table}
The left panel of Table \ref{tab:simulation_results} shows the simulation results for the uncorrected estimator. For the smallest sample size, (50, 10), the relative bias is substantial at 18.465\%, but decreases steadily with increasing sample size. This is as expected, since the theory predicts that the bias is of order $1 / N + 1 / T$ and should therefore decrease as the panel dimensions increase. For the largest sample size, (250, 50), the relative bias reduces to 3.113\%. Although the bias may seem small, it is still large relative to the dispersion of the estimator. This can be seen from the second column, which shows the ratio of bias to standard deviation. More precisely, the ratio actually increases with the sample size, i.e.\ the bias problem gets worse in relative terms as the sample size increases. This bias problem is accordingly reflected in the zero coverage rates shown in the third column. Thus, as predicted by our asymptotic theory, the uncorrected estimator for our model, \eqref{eq:uncorrected_estimator}, exhibits a more severe form of asymptotic bias problem than, for example, the uncorrected estimators in \textcite{fw2016} and \textcite{wz2021}. The right panel of Table \ref{tab:simulation_results} shows the simulation results for the bias-corrected estimator. If we compare the bias-corrected and the uncorrected estimator, we find that the former outperforms the latter in every metric in every sample. For example, even for the smallest sample size, (50, 10), the bias of 18.465\% is reduced to less than 1\% and coverage rates are improved from zero to the desired nominal level of 95\%. The same applies to all other analyzed sample sizes. Overall, the simulation experiments add numerical evidence that our asymptotic results provide a reasonable approximation for samples with sufficiently large $N$ and $T$. 

\section{Empirical example}
\label{sec:empirical_example}

In the following, we apply the uncorrected and our debiased estimator to real data, using an example from international trade.

To construct a panel data set on bilateral relationships, as described in Example \ref{ex:bilateral}, we combine two data sources. The first data source is the \textit{CEPII Gravity Database}, provided by \textcite{ccm2022}.\footnote{\url{http://www.cepii.fr/CEPII/en/bdd_modele/bdd_modele_item.asp?id=8}} This database provides information on bilateral trade flows between countries over time, from different sources such as \textit{UNSD's Comtrade}, \textit{IMF DOTS}, or \textit{CEPII's BACI}, as well as other trade cost variables that are frequently used for gravity estimation. The second data source is the \textit{Regional Trade Agreements Database}, provided by \textcite{el2008}.\footnote{\url{https://www.ewf.uni-bayreuth.de/de/forschung/RTA-daten/index.html}} This database contains additional information about regional trade agreements (RTA), allowing us to distinguish between different but not mutually exclusive types, such as customs unions (CU), free trade agreements (FTA), partial scope agreements (PSA), or economic integration agreements (EIA). Because we use trade flows from \textit{CEPII's BACI}, which are only available from 1996, and restrict ourselves to the most recent year before the COVID-19 pandemic, our final sample consists of $N = 237$ countries observed between 1996 and 2019 (i.e.\ $T = 24$ years). After removing self-trade and incomplete observations, we are left with an unbalanced panel of $n = 1{,}306{,}232$ observations.

We estimate the following binary logit model,
\begin{equation}
    \label{eq:model_empirical_example}
    \ind\{\text{trade}_{ijt} > 0\} = \ind\{x_{ijt}^{\prime} \beta + \alpha_{it} + \gamma_{jt} + \rho_{ij} > \epsilon_{ijt}\} \, ,
\end{equation}
where $\text{trade}_{ijt}$ is the trade flow from exporting country $i$ to importing country $j$ at time $t$, 
\begin{equation*}
    x_{ijt} = (\text{CU}_{ijt - 1}, \text{FTA}_{ijt - 1}, \text{PSA}_{ijt - 1}, \text{EIA}_{ijt - 1}) 
\end{equation*}
is a set of RTA-type indicator variables, $\beta = (\beta_{1}, \ldots, \beta_{4})$ are the corresponding model parameters, $\alpha_{it}$, $\gamma_{jt}$, and $\rho_{ij}$ are three sets of fixed effects accounting for different sources of unobserved heterogeneity (e.g.\ market sizes, multilateral resistance, or other time-invariant trade costs), and $\epsilon_{ijt}$ is an idiosyncratic error term. We lag the RTA-type indicator variables by one period to account for the time it takes for firms to adjust to changes in trade agreements.

Table \ref{tab:empirical_example} presents uncorrected and debiased estimation results for model \eqref{eq:model_empirical_example}. 
\begin{table}[!htbp]
\centering
\caption{Uncorrected and debiased estimation results}
\begin{threeparttable}
\begin{tabular}{lrrrrrrrr}
    \toprule
    & \multicolumn{4}{c}{uncorrected} & \multicolumn{4}{c}{debiased} \\
    \cmidrule(lr){2-5}\cmidrule(lr){6-9}    
    &CU & FTA & PSA & EIA & CU & FTA & PSA & EIA \\ 
    \midrule
    \multicolumn{9}{l}{A: Model parameters} \\
    \cmidrule(lr){1-9}
    Estimate&-0.176 & 0.360 & -0.612 & -0.172 & -0.144 & 0.315 & -0.450 & -0.130 \\ 
    Std. err.&0.098 & 0.053 & 0.158 & 0.079 & 0.098 & 0.053 & 0.159 & 0.079 \\ 
    $z$-statistic&-1.786 & 6.733 & -3.859 & -2.190 & -1.460 & 5.899 & -2.834 & -1.653 \\ 
    \midrule
    \multicolumn{9}{l}{B: Odds ratios / relative risks} \\
    \cmidrule(lr){1-9}
    Estimate&0.839 & 1.433 & 0.543 & 0.842 & 0.866 & 1.370 & 0.638 & 0.878 \\ 
    Std. err.&0.083 & 0.077 & 0.086 & 0.066 & 0.085 & 0.073 & 0.101 & 0.069 \\ 
    $z$-statistic&-1.953 & 5.654 & -5.321 & -2.390 & -1.570 & 5.061 & -3.578 & -1.765 \\  
    \bottomrule
\end{tabular}
\begin{tablenotes}
    \footnotesize
    \item\textbf{Notes:} Estimation results are obtained using data on bilateral trade between $N = 237$ countries observed for $T = 24$ years; uncorrected and debiased refer to quantities based on \eqref{eq:uncorrected_estimator} and \eqref{eq:debiased_estimator}, respectively; the upper (A) and lower panel (B) report results for the model parameters and the odds ratios, respectively; odds ratios/relative risks are computed as $\exp(\hat{\beta}_{k})$ for each $k \in \{1, \ldots, 4\}$, with standard errors obtained by the Delta theorem; the $z$-statistics in panel A and B are computed as $\hat{\beta}_{k} / \text{se}(\hat{\beta}_{k})$ and $(\exp(\hat{\beta}_{k}) - 1) / \text{se}(\exp(\hat{\beta}_{k}))$ for each $k \in \{1, \ldots, 4\}$, respectively.
\end{tablenotes}
\end{threeparttable}
\label{tab:empirical_example}
\end{table}
In addition to the model parameter estimates in panel A, we also report the odds ratios (or relative risks) in panel B. Odds ratios are calculated as $\exp(\hat{\beta}_{k})$ for each $k \in \{1, \ldots, 4\}$ and are a useful metric for interpreting the results of logit models. Unlike partial effects, which are another useful metric, odds ratios only depend on the model parameter estimates and therefore do not require further theoretical investigation. We are primarily interested in analyzing the differences between inferences drawn from the uncorrected and debiased estimators. Therefore, we also investigate the corresponding test statistics for typical two-sided hypothesis tests:  $\mathbb{H}_{0} \colon \beta_{k} = 0$ for panel A and $\mathbb{H}_{0} \colon \exp(\beta_{k}) = 1$ for panel B, for each $k \in \{1, \ldots, 4\}$. Analyzing panel A, we find that debiasing the estimates substantially reduces the magnitude of the model parameter estimates. Relative to the corresponding standard errors, the reductions range between 0.3 and 1 times the standard error. The debiasing of the estimates also results in lower test statistics. For example, the estimate for EIA becomes insignificant at the 5\% level after correcting for the bias. As the odds ratios are just a function of the estimated model parameters, the findings from panel A also carry over to panel B. For example, the uncorrected estimate suggests that forming a free trade agreement increases the probability to trade by 43.4\%. However, after debiasing the estimate, we find that the increase is reduced to 37\%, which is a 6.4 percentage point reduction. Similarly, the uncorrected estimate suggests that forming a partial scope agreement reduces the probability to trade by 45.7\%. After debiasing, the decrease is reduced to 36.2\%, which is a 9.5 percentage point reduction.

The empirical example illustrates that, although the panel data set is quite large, with $N = 237$ countries observed for $T = 24$ years, debiasing the estimates significantly impacts the results and the inferences drawn.

\section{Conclusion}
\label{sec:conclusion}

We studied the asymptotic behavior of fixed effects estimators for logit models with three additive and overlapping unobserved effects in three-dimensional panels, under asymptotic sequences where all three panel dimensions grow large. To address the asymptotic bias problem of the uncorrected estimator, we proposed a debiasing procedure. The inference problem we identify is more severe than in previous studies, highlighting the need for further research on the properties of fixed effects estimators for nonlinear models with multiple unobserved effects in multi-dimensional panels. Therefore, empirical researchers should be aware of the potential pitfalls of these estimators before using them in practice.

Several interesting topics remain for future research. For instance, our results could be extended to average partial effects and other (potentially dynamic) nonlinear models, as well as to panels with more than three dimensions. Additionally, it could be useful to derive fixed-$T$ consistent fixed effects estimators, as not every panel spans a sufficiently long time period. We plan to explore some of these topics in future work.

\clearpage

{\noindent\LARGE\textbf{Appendix}}

\appendix

\section{Notation}

Before presenting the proofs, we briefly comment on the general notation used throughout the Appendix. We consider binary logit models with linear index specification $x_{ijt}^{\prime} \beta + w_{ijt}^{\prime} \phi$, where $x$ is a $N^2 T \times K$ matrix of explanatory variables and $w = (w^{(1)}, w^{(2)}, w^{(3)})$ is a $(N^2T \times 2NT + N^2)$ matrix with rows equal to
\begin{align*}
    w_{ijt}^{(1)} =& \, (\delta_{(i,t), (1, 1)}^{(1)}, \ldots, \delta_{(i,t), (1, T)}^{(1)}, \delta_{(i,t), (2, 1)}^{(1)} \ldots, \delta_{(i,t), (N, T)}^{(1)}) \, ,  \\
    w_{ijt}^{(2)} =& \, (\delta_{(j,t), (1, 1)}^{(2)}, \ldots, \delta_{(j,t), (1, T)}^{(2)}, \delta_{(j,t), (2, 1)}^{(2)}, \ldots, \delta_{(j,t), (N, T)}^{(2)}) \, ,  \\
    w_{ijt}^{(3)} =& \, (\delta_{(i,j), (1, 1)}^{(3)}, \ldots, \delta_{(i,j), (1, N)}^{(3)}, \delta_{(i,j), (2, 1)}^{(3)}, \ldots, \delta_{(i,j), (N, N)}^{(3)}) \, , 
\end{align*}
where $\delta_{(i,t), (i^{\prime}, t^{\prime})}^{(1)} = \ind\{i = i^{\prime}\}\ind\{t = t^{\prime}\}$, $\delta_{(j,t), (j^{\prime}, t^{\prime})}^{(2)} = \ind\{j = j^{\prime}\}\ind\{t = t^{\prime}\}$, and $\delta_{(i,j), (i^{\prime}, j^{\prime})}^{(3)} = \ind\{i = i^{\prime}\}\ind\{j = j^{\prime}\}$. The model parameters are $\beta$ and $\phi$ are nuisance parameters. We use $\mu^{\langle 1 \rangle}(z) = \partial_{z} \mu(z)$, $\mu^{\langle 2 \rangle}(z) = \partial_{z^2} \mu(z)$, and $\mu^{\langle 3 \rangle}(z) = \partial_{z^3} \mu(z)$ to denote the first-, second-, and third-order derivatives of the logistic cumulative distribution function $\mu(z) = (1 + \exp(- z))^{- 1}$. To make the notation more concise, we often write $\mu_{ijt}$ instead of $\mu(x_{ijt}^{\prime} \beta^{0} + w_{ijt}^{\prime} \phi^{0})$. Similarly, we write $\hat{\mu}_{ijt} = \mu(x_{ijt}^{\prime} \hat{\beta} + w_{ijt}^{\prime} \hat{\phi})$ and or $\check{\mu}_{ijt} = \mu(x_{ijt}^{\prime} \check{\beta} + w_{ijt}^{\prime} \check{\phi})$. Other quantities, like $\mu_{ijt}^{\langle 1 \rangle} = \mu^{\langle 1 \rangle}(x_{ijt}^{\prime} \beta^{0} + w_{ijt}^{\prime} \phi^{0})$, are defined accordingly. Further, we use $e_{g}$ to denote basis vectors with a one in the $g$-th coordinate and zeros everywhere else, and define $u(\beta, \phi) = y - \mu(x \beta + w \phi)$.

Our asymptotic expansions are based on a projection approach, following \textcite{fw2016}. Therefore, we need to define the corresponding projections. Let $\partial_{\phi \phi^{\prime}} L(\beta, \phi) =  H(\beta, \phi) = (w^{\prime} \mu^{\langle 1 \rangle}(x \beta + w \phi) w + c_{1} v v^{\prime}) / \sqrt{NT}$ denote the incidental parameter Hessian. Then, $Q(\beta, \phi) = (w (H(\beta, \phi))^{- 1} w^{\prime} \diag(\mu^{\langle 1 \rangle}(x \beta + w \phi))) / \sqrt{NT}$ is a projection matrix, sometimes called hat maker matrix, that maps a vector to a vector of fitted values. This is the projection matrix of a weighted least squares problem where $\mu^{\langle 1 \rangle}(x \beta + w \phi)$ are the weights. Additionally, we define $M(\beta, \phi) = \eye_{N^2T} - Q(\beta, \phi)$, which is sometimes called residual maker matrix as it maps a vector to a vector of residuals. Further, as it arises in the expansions, we also define the linear operator $\mathbb{Q}(\beta, \phi) = (w (H(\beta, \phi))^{- 1} w^{\prime}) / \sqrt{NT}$. All asymptotic statements are based on sequences where $N, T \rightarrow \infty$.

For vectors, we use the $p$-norm, which is defined as $\norm{z}_{p} = (\sum_{n = 1}^{\dim(z)} \lvert z_{n} \rvert^{p})^{1 / p}$ for a vector $z$. For example, the $p$-norm for $p = 2$ is the Euclidean norm, the $p$-norm for $p = 1$ is the taxicab norm, and the $p$-norm for $p = \infty$ is the maximum norm. For matrices, we use norms induced by the vector $p$-norms. For example, the spectral norm, column-sum norm, and row-sum norm are all induced by the $p$-norm for $p = 2$, $p = 1$, and $p = \infty$, respectively. Finally, the max norm of a matrix $A$ is defined as $\norm{A}_{\max} = \max_{m, n} \, \lvert A_{m, n} \rvert$, where $A_{m, n}$ refers to the element in the $m$-th row and $n$-th column.

\section{Proofs of main text results}

\subsection{Proof of Theorem \ref{theorem:uncorrected}}
\label{app:theorem_uncorrected}

By Lemma \ref{lemma:second_order_expansion}, we have
\begin{equation*}
    N\sqrt{T} (\hat{\beta} - \beta^{0}) = W^{- 1} U^{(0)} + W^{- 1} U^{(1)} + o_{p}(1) \, ,
\end{equation*}
where 
\begin{align*}
    W =& \, \frac{1}{N^2T} (M x)^{\prime} \diag(\mu^{\langle 1 \rangle}) x \, ,\\
    U^{(0)} =& \, \frac{1}{N \sqrt{T}} (M x)^{\prime} u \, ,\\
    U^{(1)} =& \, - \frac{1}{2 N \sqrt{T}} (\mathbb{Q} u)^{\prime} \diag(\mu^{\langle 2 \rangle} \odot \mathbb{Q} u)  M x \, .
\end{align*}
Further, we have that $U^{(0)} \overset{d}{\rightarrow} \mathcal{N}(0, \overline{W})$, where $\overline{W} = \LEX{W}$. To finish the proof, we need to bound the $K$-dimensional vector $U^{(1)}$.

First note that
\begin{equation*}
    \norm{U^{(1)}}_{2} \leq \norm{U^{(1)}}_{1} = \sum_{k = 1}^{K} \big\lvert U_{k}^{(1)} \big\rvert \, ,
\end{equation*}
where $U_{k}^{(1)}$ denotes the $k$-th element in $U^{(1)}$. Then,
\begin{align*}
    U_{k}^{(1)} =& \, \frac{1}{2 N \sqrt{T}} \big\lvert (\mathbb{Q} u)^{\prime} \diag(\mu^{\langle 2 \rangle} \odot M x e_{k}) \mathbb{Q} u \big\rvert \\
    \leq& \, \frac{1}{N \sqrt{T}} \big\lvert (\mathbb{Q} u)^{\prime} \mathbb{Q} u \big\rvert \norm{\mu^{\langle 2 \rangle}}_{\infty} \norm{M x}_{\max} \\
    =& \, \mathcal{O}_{P}((NT)^{1 / 4}) \, ,
\end{align*}
where we used that $(\mathbb{Q} u)^{\prime} \mathbb{Q} u = \mathcal{O}_{P}(NT)$ by Lemma \ref{lemma:bound_second_order} i), and $\norm{\mu^{\langle 2 \rangle}}_{\infty} = \mathcal{O}_{P}(1)$ and $\norm{M x}_{\max} = \mathcal{O}_{P}(1)$ by Assumption 1 iii).

Thus, we conclude, since $U^{(0)} \overset{d}{\rightarrow} \mathcal{N}(0, \overline{W})$, for the asymptotic distribution to have a finite expected value, both sides would have to be divided by $(NT)^{1 / 4}$. However, this would cause the variance $\diag(\overline{W})$ to decrease towards zero. Thus, it is not possible to balance the order of the bias and the variance, i.e.\ the uncorrected maximum likelihood estimator has a degenerating asymptotic distribution.\hfill\qedsymbol

\subsection{Proof of Theorem \ref{theorem:debiased}}
\label{app:theorem_debiased}

The beginning of the proof is the same as for Theorem 1. Thus, by Lemma \ref{lemma:second_order_expansion}, we have
\begin{equation}
    \label{eq:second_order_expansion}
    N\sqrt{T} (\hat{\beta} - \beta^{0}) = W^{- 1} U^{(0)} + W^{- 1} U^{(1)} + o_{p}(1) \, ,
\end{equation}
where 
\begin{align*}
    W =& \, \frac{1}{N^2T} (M x)^{\prime} \diag(\mu^{\langle 1 \rangle}) x \, ,\\
    U^{(0)} =& \, \frac{1}{N \sqrt{T}} (M x)^{\prime} u \overset{d}{\rightarrow} \mathcal{N}(0, \overline{W}) \, ,\\
    U^{(1)} =& \, - \frac{1}{2 N \sqrt{T}} (\mathbb{Q} u)^{\prime} \diag(\mu^{\langle 2 \rangle} \odot \mathbb{Q} u)  M x \, .
\end{align*}
Rearranging \eqref{eq:second_order_expansion} yields
\begin{equation*}
    N \sqrt{T} \, (\hat{\beta} - \beta^{0}) - W^{- 1} U^{(1)} = W^{- 1} U^{(0)} + o_{p}(1) \, .
\end{equation*}
In the following, we want to show that 
\begin{equation*}
    U^{(1)} / (NT)^{1 / 4} \overset{p}{\rightarrow} B_{\alpha} + B_{\gamma} + B_{\rho} = B \, ,
\end{equation*}
where
\begin{align*}
     B_{\alpha} =& \, - \frac{1}{2 NT} \sum_{i = 1}^{N} \sum_{t = 1}^{T} \frac{\sum_{j = 1}^{N} \mu_{ijt}^{\langle 2 \rangle} \tilde{x}_{ijt}}{\sum_{j = 1}^{N} \mu_{ijt}^{\langle 1 \rangle}} \, , \\
     B_{\gamma} =& \, - \frac{1}{2 NT} \sum_{j = 1}^{N} \sum_{t = 1}^{T} \frac{\sum_{i = 1}^{N} \mu_{ijt}^{\langle 2 \rangle} \tilde{x}_{ijt}}{\sum_{i = 1}^{N} \mu_{ijt}^{\langle 1 \rangle}} \, , \\
     B_{\rho} =& \, - \frac{1}{2 N^2} \sum_{i = 1}^{N} \sum_{j = 1}^{N} \frac{\sum_{t = 1}^{T} \mu_{ijt}^{\langle 2 \rangle} \tilde{x}_{ijt}}{\sum_{t = 1}^{T} \mu_{ijt}^{\langle 1 \rangle}} \, ,
\end{align*}
are the bias components defined in Section \ref{sec:asymptotic_theory}, and we use $\tilde{x} = M x$ to simplify the notation. Note that, by conditional independence and Assumption iii), we already have $W \overset{p}{\rightarrow} \overline{W}$. 

Using that $\mathbb{Q} = \mathbb{Q}_{\phi} + \mathbb{Q}_{R}$, we can decompose $U^{(1)}$ into
\begin{align*}
    U^{(1)} =& \, (\tilde{x}^{\prime} \diag(\mu^{\prime \prime} \odot \mathbb{Q} u) \mathbb{Q} u) / (2 N \sqrt{T}) \\
    =& \, (\tilde{x}^{\prime} \diag(\mu^{\prime \prime} \odot \mathbb{Q}_{\phi} u) \mathbb{Q}_{\phi} u) / (2 N \sqrt{T}) + \\
    & \, (\tilde{x}^{\prime} \diag(\mu^{\prime \prime} \odot \mathbb{Q}_{R} u) \mathbb{Q}_{\phi} u) / (N \sqrt{T}) + \\
    & \, (\tilde{x}^{\prime} \diag(\mu^{\prime \prime} \odot \mathbb{Q}_{R} u) \mathbb{Q}_{R} u) / (2 N \sqrt{T}) \, ,
\end{align*}
so that
\begin{align*}
    \norm{U^{(1)} / (NT)^{1 / 4} - B}_{2} \leq& \, \norm{(\tilde{x}^{\prime} \diag(\mu^{\prime \prime} \odot \mathbb{Q}_{\phi} u) \mathbb{Q}_{\phi} u) / (2 N T) - (B_{\alpha} + B_{\gamma} + B_{\rho})}_{2} + \\
    & \, \norm{(\tilde{x}^{\prime} \diag(\mu^{\prime \prime} \odot \mathbb{Q}_{R} u) (2 \mathbb{Q}_{\phi} u + \mathbb{Q}_{R} u)) / (2 N T)}_{2} \\
    =& \, \norm{C_{\phi\phi}}_{2} + \norm{C_{R}}_{2}\, .
\end{align*}
by the triangle inequality. 

Further, using that $\mathbb{Q}_{\phi} = \mathbb{Q}_{\alpha} + \mathbb{Q}_{\gamma} + \mathbb{Q}_{\rho}$ along with the triangle inequality, $\norm{C_{\phi\phi}}_{2}$ can be further decomposed into
\begin{align*}
    \norm{C_{\phi\phi}}_{2} \leq& \, \norm{(\tilde{x}^{\prime} \diag(\mu^{\prime \prime} \odot \mathbb{Q}_{\alpha} u) \mathbb{Q}_{\alpha} u) / (2 N T) - B_{\alpha}}_{2} + \\
    & \, \norm{(\tilde{x}^{\prime} \diag(\mu^{\prime \prime} \odot \mathbb{Q}_{\gamma} u) \mathbb{Q}_{\gamma} u) / (2 N T) - B_{\gamma}}_{2} + \\
    & \, \norm{(\tilde{x}^{\prime} \diag(\mu^{\prime \prime} \odot \mathbb{Q}_{\rho} u) \mathbb{Q}_{\rho} u) / (2 N T) - B_{\rho}}_{2} + \\
    & \, \norm{(\tilde{x}^{\prime} \diag(\mu^{\prime \prime} \odot \mathbb{Q}_{\alpha} u) \mathbb{Q}_{\gamma} u) / (N T)}_{2} + \\
    & \, \norm{(\tilde{x}^{\prime} \diag(\mu^{\prime \prime} \odot \mathbb{Q}_{\alpha} u) \mathbb{Q}_{\rho} u) / (N T)}_{2} + \\
    & \, \norm{\big(\tilde{x}^{\prime} \diag(\mu^{\prime \prime} \odot \mathbb{Q}_{\gamma} u) \mathbb{Q}_{\rho} u) / (N T)}_{2} \\
    =& \, \norm{C_{\alpha\alpha}}_{2} + \norm{C_{\gamma\gamma}}_{2} + \norm{C_{\rho\rho}}_{2} + \norm{C_{\alpha\gamma}}_{2} + \norm{C_{\alpha\rho}}_{2} + \norm{C_{\gamma\rho}}_{2} \, .
\end{align*}

We start with the bounds for the first three components of $\norm{C_{\phi\phi}}_{2}$. Let
\begin{equation*}
    C_{\alpha\alpha} = \frac{1}{2 NT} \sum_{i = 1}^{N} \sum_{t = 1}^{T} \frac{(\sum_{j = 1}^{N} \mu_{ijt}^{\langle 2 \rangle} \tilde{x}_{ijt}) ((\sum_{j = 1}^{N} \mu_{ijt}^{\langle 1 \rangle}) - (\sum_{j = 1}^{N} u_{ijt})^{2})}{(\sum_{j = 1}^{N} \mu_{ijt}^{\langle 1 \rangle})^2} \, ,
\end{equation*}
so that
\begin{equation*}
    \norm{C_{\alpha\alpha}}_{2} \leq \frac{K}{2 N^{2} T} \norm{\tilde{x}}_{\max} \, \norm{\mu^{\langle 2 \rangle}}_{\infty} \, \norm{(\mu^{\langle 1 \rangle})^{- 1}}_{\infty}^{2} \, \Big\lvert \sum_{i = 1}^{N} \sum_{t = 1}^{T} \big(\big(\sum_{j = 1}^{N} \mu_{ijt}^{\langle 1 \rangle}\big) - \big(\sum_{j = 1}^{N} u_{ijt}\big)^{2}\big) \Big\rvert \, .
\end{equation*}
Further, let $\xi_{it} = (\sum_{j = 1}^{N} \mu_{ijt}^{\langle 1 \rangle}) - (\sum_{j = 1}^{N} u_{ijt})^{2}$, where
\begin{equation*}
    \EX{\xi_{it}} = \EX{\CEX{\xi_{it}}} = \EX{\big(\sum_{j = 1}^{N} \mu_{ijt}^{\langle 1 \rangle}\big)} - \EX{\CEX{\big(\sum_{j = 1}^{N} u_{ijt}\big)^{2}}} = 0 
\end{equation*}
as $\CEX{(u_{ijt})^2} = \mu_{ijt} (1 - \mu_{ijt}) = \mu_{ijt}^{\langle 1 \rangle}$ is implied by the logistic distribution function. By conditional independence and Assumption iii), it follows that
\begin{equation*}
    \EX{\Big( \sum_{i = 1}^{N} \sum_{t = 1}^{T} \xi_{it} \Big)^2} = \mathcal{O}(NT) \, ,
\end{equation*}
which implies $\sum_{i = 1}^{N} \sum_{t = 1}^{T} \xi_{it} = \mathcal{O}_{P}(\sqrt{NT})$. Thus, using that $\norm{(\mu^{\langle 1 \rangle})^{- 1}}_{\infty} = \mathcal{O}_{P}(1)$, $\norm{\mu^{\langle 2 \rangle}}_{\infty} = \mathcal{O}_{P}(1)$, and $\norm{\tilde{x}}_{\max} = \mathcal{O}_{P}(1)$, we conclude that $\norm{C_{\alpha\alpha}}_{2} = o_{P}(1)$. The bounds $\norm{C_{\gamma\gamma}}_{2} = o_{P}(1)$ and $\norm{C_{\rho\rho}}_{2} = o_{P}(1)$ follow analogously.

Next, we bound the last three components of $\norm{C_{\phi\phi}}_{2}$. Let
\begin{equation*}
    C_{\alpha\gamma} = \frac{1}{NT} \sum_{i = 1}^{N} \sum_{j = 1}^{N} \sum_{t = 1}^{T} \frac{(\mu_{ijt}^{\langle 2 \rangle} \tilde{x}_{ijt}) (\sum_{i^{\prime} = 1}^{N} u_{i^{\prime}jt}) (\sum_{j^{\prime} = 1}^{N} u_{ij^{\prime}t})}{(\sum_{i^{\prime} = 1}^{N} \mu_{i^{\prime}jt}^{\prime}) (\sum_{j^{\prime} = 1}^{N} \mu_{ij^{\prime}t}^{\prime})} \, ,
\end{equation*}
so that
\begin{equation*}
    \norm{C_{\alpha\gamma}}_{2} \leq \frac{K}{N^3 T} \norm{\tilde{x}}_{\max} \, \norm{\mu^{\langle 2 \rangle}}_{\max} \, \norm{(\mu^{\langle 1 \rangle})^{- 1}}_{\max}^{2} \, \Big\lvert \sum_{i = 1}^{N} \sum_{j = 1}^{N} \sum_{t = 1}^{T} \big(\sum_{i^{\prime} = 1}^{N} u_{i^{\prime}jt}\big) \big(\sum_{j^{\prime} = 1}^{N} u_{ij^{\prime}t}\big) \Big\rvert \, .
\end{equation*}
Further, by conditional independence and the fact that $\norm{u}_{\infty} \leq 1$, we get
\begin{equation*}
    \EX{\CEX{\Big( \sum_{i = 1}^{N} \sum_{j = 1}^{N} \sum_{t = 1}^{T} \big(\sum_{i^{\prime} = 1}^{N} u_{i^{\prime}jt}\big) \big(\sum_{j^{\prime} = 1}^{N} u_{ij^{\prime}t}\big) \Big)^{2}}} = \mathcal{O}_{P}(N^{4} T^{2}) \, ,
\end{equation*}
which implies $\sum_{i = 1}^{N} \sum_{j = 1}^{N} \sum_{t = 1}^{T} (\sum_{i^{\prime} = 1}^{N} u_{i^{\prime}jt}) (\sum_{j^{\prime} = 1}^{N} u_{ij^{\prime}t}) = \mathcal{O}_{P}(N^2T)$. Thus, using that $\norm{(\mu^{\langle 1 \rangle})^{- 1}}_{\infty} = \mathcal{O}_{P}(1)$, $\norm{\mu^{\langle 2 \rangle}}_{\infty} = \mathcal{O}_{P}(1)$, and $\norm{\tilde{x}}_{\max} = \mathcal{O}_{P}(1)$, we conclude that $\norm{C_{\alpha\gamma}}_{2} = o_{P}(1)$. The bounds $\norm{C_{\alpha\rho}}_{2} = o_{P}(1)$ and $\norm{C_{\gamma\rho}}_{2} = o_{P}(1)$ follow analogously.

Finally,
\begin{align*}
    \norm{C_{R}}_{2} \leq& \, \sqrt{NT} \norm{\tilde{x}}_{\max} \norm{\mu^{\prime \prime}}_{\infty} \norm{\mathbb{Q}_{R} u}_{\infty} (2 \norm{\mathbb{Q}_{\phi} u}_{\infty} + \norm{\mathbb{Q}_{R} u}_{\infty})  \\
    =& \, \mathcal{O}_{P}((NT)^{- 1 / 2}) + \mathcal{O}_{P}((NT)^{- 1}) = o_{P}(1) \, ,
\end{align*}
where $\norm{\tilde{x}}_{\max} = \mathcal{O}_{P}(1)$ and $\norm{\mu^{\prime \prime}}_{\max} = \mathcal{O}_{P}(1)$ by Assumption 1 iii) and, $\norm{\mathbb{Q}_{R} u}_{\infty} = \mathcal{O}_{P}((NT)^{- 3 / 4})$ and $\norm{\mathbb{Q}_{\phi} u}_{\infty} = \mathcal{O}_{P}((NT)^{- 1 / 4})$ by Lemma \ref{lemma:linear_operator}.

Bringing all components together, we get 
\begin{equation*}
    \norm{U^{(1)} / (NT)^{1 / 4} - (B_{\alpha} + B_{\gamma} + B_{\rho})}_{2} = o_{P}(1) \, ,
\end{equation*}
which implies that $U^{(1)} / (NT)^{1 / 4} \overset{p}{\rightarrow} B$. Furthermore, using Slutsky's theorem, we have $W^{- 1} U^{(1)} / (NT)^{1 / 4} \overset{p}{\rightarrow} \overline{W}^{- 1} (B_{\alpha} + B_{\gamma} + B_{\rho}) = b$ and obtain
\begin{equation*}
    N\sqrt{T} (\hat{\beta} - \beta^{0} - b / \sqrt{NT}) \overset{d}{\rightarrow} \mathcal{N}(0, \overline{W}^{- 1}) \, .
\end{equation*}
Thus, we conclude that the bias-corrected estimator has a non-degenerate asymptotic distribution.\hfill\qedsymbol

\subsection{Proof of Lemma \ref{lemma:consistency_bias_variance}}
\label{app:lemma_consistency_bias_variance}

Given that $\norm{\hat{\beta} - \beta^{0}}_{2} \overset{p}{\rightarrow} 0$ and $\norm{\hat{\phi}(\hat{\beta}) - \phi^{0}}_{\infty} \overset{p}{\rightarrow} 0$ by Lemma \ref{lemma:convergence_rates}, consistency of the estimators for bias variance components follows immediately from Lemma S.1 in \textcite{fw2016} or from Lemma 7 in \textcite{cfw2020}. \hfill\qedsymbol 

\section{Intermediate results}

\subsection{Incidental parameter Hessian}
\label{app:ip_hessian}

We need to ensure that the incidental parameter Hessian,
\begin{equation*}
    \frac{\partial^{2} L(\beta, \phi)}{\partial \phi \partial \phi^{\prime}} = H(\beta, \phi) = (w^{\prime} \diag(\mu^{\langle 1 \rangle}(x \beta + w \phi)) w + c_{1} v v^{\prime}) / \sqrt{NT} \, ,
\end{equation*}
is asymptotically positive definite, for all $c_{1} > 0$. Therefore, we need to impose specific constraints on the incidental parameters $\phi = (\alpha, \gamma, \rho)$. It is important that the constraints $v^{\prime} \phi$ enter the log-likelihood function as a quadratic ``penalty term''. This ensures that $v v^{\prime}$ appears in the Hessian, which can be used to solve the rank deficiency problem (due to perfect collinearity) in $w$.

Since we use $N = I = J$ in our analysis, \eqref{eq:constraints} becomes
\begin{align*}
	v & = 	
	\begin{pmatrix}
        1_{N} \otimes \eye_{T}   & \eye_{N} \otimes 1_{T}     & 0_{NT \times N}	\\
        - 1_{N} \otimes \eye_{T} & 0_{NT \times N}         & \eye_{N} \otimes 1_{T} \\
        0_{N^2 \times T}      & - \eye_{N} \otimes 1_{N}   & - 1_{N} \otimes \eye_{N}
	\end{pmatrix} = 
	\begin{pmatrix}
		v_{1} & 	v_{2} & 	v_{3}\\
		v_{4} & 	v_{5} & 	v_{6}\\
		v_{7} & 	v_{8} & 	v_{9}
	\end{pmatrix} \, .
\end{align*}
It is important to note that the sparsity pattern of the incidental parameter Hessian, as described in Section \ref{sec:asymptotic_theory}, remains asymptotically identical after imposing the constraints.

Lemma \ref{lemma:ip_hessian} shows that Assumption 1 ensures that the smallest eigenvalue of the incidental parameter Hessian is asymptotically bounded away from zero by a positive constant.
\begin{lemma}[Incidental parameter Hessian]\label{lemma:ip_hessian}
    Let Assumption 1 hold. Then,
    \begin{equation*}
        \underset{N, T \rightarrow \infty}{\lim} \lambda_{\min}(H(\beta, \phi)) \geq \frac{3}{7} \, c_{\min} > 0 \, ,
    \end{equation*}
    where $\lambda_{\min}(\cdot)$ is the smallest eigenvalue of a matrix and $0 < c_{\min} \leq \mu^{\langle 1 \rangle}(\beta, \phi)$. Moreover, $\norm{(H(\beta, \phi))^{- 1}}_{2} = \mathcal{O}_{P}(1)$.
\end{lemma}

\noindent\textbf{Proof of Lemma \ref{lemma:ip_hessian}.} By the Courant–Fischer–Weyl min-max principle, we have
\begin{align*}
    \lambda_{\min}(H(\beta, \phi)) =& \, \min_{\{\Delta \in \mathbb{R}^{2 NT + N^2} \colon \norm{\Delta} = 1\}} \Delta^{\prime}((w^{\prime} \diag(\mu^{\langle 1 \rangle}(x \beta + w \phi)) w + c_{1} v v^{\prime}) / \sqrt{NT})\Delta \\
    \geq& \, c_{\min} \, \min_{\{\Delta \in \mathbb{R}^{2 NT + N^2} \colon \norm{\Delta} = 1\}} \Delta^{\prime}((w^{\prime} w + (c_{1} / c_{\min}) v v^{\prime}) / \sqrt{NT})\Delta \\
    =& \, c_{\min} \, \lambda_{\min}((w^{\prime} w + (c_{1} / c_{\min}) v v^{\prime}) / \sqrt{NT}) \, ,
\end{align*}
where we used that $0 < c_{\min} \leq \mu^{\langle 1 \rangle}(\beta, \phi) \leq c_{\max} < \infty$ is implied by Assumption 1 iii). Further, choosing $c_{1} = c_{\min}$ and taking the limit, we get
\begin{equation*}
    \underset{N, T \rightarrow \infty}{\lim} \lambda_{\min}(H(\beta, \phi)) \geq c_{\min} \, \underset{N, T \rightarrow \infty}{\lim} \, \lambda_{\min}(\mathcal{H}) \, ,
\end{equation*}
where $\mathcal{H} = (w^{\prime} w + v v^{\prime}) / \sqrt{NT}$ is just the incidental parameter Hessian for linear three-dimensional panel models that use the same linear index specification as we do. In the following, we use a two-part strategy to show that $\lambda_{\min}(\mathcal{H})$ is uniformly bounded away from zero by constant. First, we exploit the block structure of $\mathcal{H}$ to compute $\mathcal{H}^{- 1}$. Second, we use that $(\lambda_{\min}(\mathcal{H}))^{- 1} = \norm{\mathcal{H}^{- 1}}_{2} \leq \norm{\mathcal{H}^{- 1}}_{\infty}$ to bound $\lambda_{\min}(\mathcal{H})$.

We start with the computation of $\mathcal{H}^{- 1}$. Note that
\begin{equation*}
    \mathcal{H} = \begin{pmatrix}
            A & B^{\prime} & D^{\prime} \\
            B & C & E^{\prime} \\
            D & E & F
    \end{pmatrix}
\end{equation*}
has the following $3 \times 3$ block structure, where
\begin{align*}
    A =& \, ((w^{(1)})^{\prime} w^{(1)} + v_{1} v_{1}^{\prime} + v_{2} v_{2}^{\prime}) / \sqrt{NT} \, , \\
    B =& \, ((w^{(2)})^{\prime} w^{(1)} + v_{4} v_{1}^{\prime}) / \sqrt{NT} \, , \\
    C =& \, ((w^{(2)})^{\prime} w^{(2)} + v_{4} v_{4}^{\prime} + v_{6} v_{6}^{\prime}) / \sqrt{NT} \, , \\
    D =& \, ((w^{(3)})^{\prime} w^{(1)} + v_{8} v_{2}^{\prime}) / \sqrt{NT} \, , \\
    E =& \, ((w^{(3)})^{\prime} w^{(2)} + v_{9} v_{6}^{\prime}) / \sqrt{NT} \, , \\
    F =& \, ((w^{(3)})^{\prime} w^{(3)} + v_{8} v_{8}^{\prime} + v_{9} v_{9}^{\prime}) / \sqrt{NT} \, .
\end{align*}
As formulas for block matrix inversion are only available for $2 \times 2$ block matrices, we reformulate $\mathcal{H}$ as $2 \times 2$ block matrix,
\begin{equation*}
    \mathcal{H} = \begin{pmatrix}
            G & P^{\prime} \\
            P & F
    \end{pmatrix} \, ,
\end{equation*}
where
\begin{equation*}
    G = \begin{pmatrix}
            A & B^{\prime} \\
            B & C 
        \end{pmatrix} \quad \text{and} \quad 
    P = \begin{pmatrix}
            D & E
        \end{pmatrix} \, .
\end{equation*}
Using block matrix inversion, it follows that
\begin{equation*}
    \mathcal{H}^{- 1} = \begin{pmatrix}
            (G - P^{\prime} F^{- 1} P)^{- 1} & - (G - P^{\prime} F^{- 1} P)^{- 1} P^{\prime} F^{- 1} \\
            - F^{- 1} P (G - P^{\prime} F^{- 1} P)^{- 1} & F^{- 1} + F^{- 1} P (G - P^{\prime} F^{- 1} P)^{- 1} P^{\prime} F^{- 1}
    \end{pmatrix}
\end{equation*}
and 
\begin{equation*}
    G^{- 1} = \begin{pmatrix}
            (A - B^{\prime} C^{- 1} B)^{- 1} & - (A - B^{\prime} C^{- 1} B)^{- 1} B^{\prime} C^{- 1} \\
            - C^{- 1} B (A - B^{\prime} C^{- 1} B)^{- 1} & C^{- 1} + C^{- 1} B (A - B^{\prime} C^{- 1} B)^{- 1} B^{\prime} C^{- 1}
    \end{pmatrix} \, .
\end{equation*}
Further, since
\begin{align*}
    (G - P^{\prime} F^{- 1} P)^{- 1} =& \, G^{- 1} (\eye_{2NT} - G^{- 1} P^{\prime} F^{- 1} P)^{- 1} \, , \\
    (A - B^{\prime} C^{- 1} B)^{- 1} =& \, A^{- 1} (\eye_{NT} - A^{- 1} B^{\prime} C^{- 1} B)^{- 1} \, ,
\end{align*}
it turns out that the computation of $\mathcal{H}^{- 1}$ essentially depends on $A^{- 1}$, $C^{- 1}$, and $F^{- 1}$. All three matrices, $A$, $C$, and $F$, have a similar structure that can be exploited to compute their inverses. For example,
\begin{equation*}
    A = ((w^{(1)})^{\prime} w^{(1)} + v_{2} v_{2}^{\prime}) / \sqrt{NT} + v_{1} v_{1}^{\prime} / \sqrt{NT} = M + v_{1} v_{1}^{\prime} / \sqrt{NT} \, , 
\end{equation*}
where
\begin{equation*}
    M = \underbrace{\diag(\eye_{T} + 1_{T} 1_{T}^{\prime} / \sqrt{N T}, \ldots, \eye_{T} + 1_{T} 1_{T}^{\prime} / \sqrt{N T})}_{N \, \text{blocks}}
\end{equation*}
is a block diagonal matrix with blocks that can be inverted using the Sherman-Morrison rank-one update formula, i.e.\
\begin{equation*}
    M^{- 1} = \underbrace{\diag(\eye_{T} - 1_{T} 1_{T}^{\prime} / (2 \sqrt{N T}), \ldots, \eye_{T} - 1_{T} 1_{T}^{\prime} / (2 \sqrt{N T}))}_{N \, \text{blocks}} \, .
\end{equation*}
Further, using the Woodbury identity, we get
\begin{equation*}
    A^{- 1} =  M^{- 1} - M^{- 1} v_{1} (\eye_{T} + v_{1}^{\prime} M^{- 1} v_{1} / \sqrt{NT})^{- 1} v_{1}^{\prime} M^{- 1} / \sqrt{NT} \, ,
\end{equation*}
where
\begin{equation*}
    \eye_{T} + v_{1}^{\prime} M^{- 1} v_{1} / \sqrt{NT} =  2 \, \eye_{T} - 1_{T} 1_{T}^{\prime} / (2 \sqrt{N T}) \, .
\end{equation*}
Thus, we can use again the Sherman-Morrison rank-one update formula and get
\begin{equation*}
    A^{- 1} = M^{- 1} - \begin{pmatrix}
        - \big(\frac{1}{2 \sqrt{NT}}\big) \eye_{T} + \big(\frac{1}{3 NT}\big) 1_{T} 1_{T}^{\prime} & \cdots & - \big(\frac{1}{2 \sqrt{NT}}\big) \eye_{T} + \big(\frac{1}{3 NT}\big) 1_{T} 1_{T}^{\prime} \\
        \vdots & \ddots & \vdots \\
        - \big(\frac{1}{2 \sqrt{NT}}\big) \eye_{T} + \big(\frac{1}{3 NT}\big) 1_{T} 1_{T}^{\prime} & \cdots & - \big(\frac{1}{2 \sqrt{NT}}\big) \eye_{T} + \big(\frac{1}{3 NT}\big) 1_{T} 1_{T}^{\prime}
    \end{pmatrix} \, .
\end{equation*}
The other inverses, $C^{- 1}$ and $F^{- 1}$, can be computed analogously. Further, since we use $I = J = N$ in our analysis, it follows that $C^{- 1} = A^{- 1}$. Finally, we use that $F^{- 1} P = 0$ and $C^{- 1} B = 0$, and find that
\begin{equation*}
    \mathcal{H}^{- 1} = \begin{pmatrix}
            A^{- 1} & 0 & 0 \\
            0 & C^{- 1} & 0 \\
            0 & 0 & F^{- 1} \\
    \end{pmatrix}
\end{equation*}
is a block diagonal matrix.

For the second part, we use that
\begin{equation*}
    \norm{\mathcal{H}^{- 1}}_{2} \leq \norm{\mathcal{H}^{- 1}}_{\infty} = \max(\norm{A^{- 1}}_{\infty}, \norm{C^{- 1}}_{\infty}, \norm{F^{- 1}}_{\infty}) = \max(\norm{A^{- 1}}_{\infty}, \norm{F^{- 1}}_{\infty}) \, ,
\end{equation*}
where the last equality follows from the fact that $A^{- 1} = C^{- 1}$ in our analysis. Using the triangle inequality, we get
\begin{align*}
    \norm{A^{- 1}}_{\infty} =& \, \Big\lvert 1 - \frac{1}{\sqrt{NT}} + \frac{1}{3 NT} \Big\rvert + (N + T - 2) \Big\lvert - \frac{1}{2\sqrt{NT}} + \frac{1}{3 NT} \Big\rvert + \\
    & \, (N - 1) (T - 1) \Big\lvert \frac{1}{3 NT} \Big\rvert \\
    \leq& \, \frac{4}{3} + \frac{N + T}{2\sqrt{NT}} 
\end{align*}
and 
\begin{align*}
    \norm{F^{- 1}}_{\infty} =& \, \Big\lvert 1 - \frac{1}{\sqrt{NT}} + \frac{1}{3 NT} \Big\rvert + (2 N - 2) \Big\lvert - \frac{1}{2\sqrt{NT}} + \frac{1}{3 NT} \Big\rvert + (N - 1)^2 \Big\lvert \frac{1}{3 NT} \Big\rvert \\
    \leq& \, 1 + \frac{N}{\sqrt{NT}} + \frac{N^2}{3 NT} \, .
\end{align*}
Taking limits and using that $N \sim \mathcal{O}(T)$ given Assumption v), we get
\begin{equation*}
    \underset{N, T \rightarrow \infty}{\lim} \, (\lambda_{\min}(\mathcal{H}))^{- 1} \leq \underset{N, T \rightarrow \infty}{\lim} \, \max(\norm{A^{- 1}}_{\infty}, \norm{F^{- 1}}_{\infty}) = \frac{7}{3} \, .
\end{equation*}
Finally, we conclude that
\begin{equation*}
    \underset{N, T \rightarrow \infty}{\lim} \lambda_{\min}(H(\beta, \phi)) \geq \frac{3}{7} \, c_{\min} > 0 \, , 
\end{equation*}
i.e.\ $H(\beta, \phi)$ is positive definite and $\norm{H(\beta, \phi)}_{2} = \mathcal{O}_{P}(1)$.\hfill\qedsymbol

Before presenting Lemma \ref{lemma:inverse_ip_hessian}, which shows that the inverse of the incidental parameter Hessian matrix can be asymptotically regarded as (weakly) diagonal dominant matrix, note that the incidental parameter Hessian has a specific block structure and can be decomposed into a diagonal matrix plus remainder matrix,
\begin{equation*}
    H(\beta, \phi) = D(\beta, \phi) + G(\beta, \phi) \, ,
\end{equation*}
where
\begin{align}
    D(\beta, \phi)  =& \, \begin{pmatrix}
            D^{(1)}(\beta, \phi) & 0_{NT \times NT} & 0_{N^{2} \times N^{2}} \\
            0_{NT \times NT} & D^{(2)}(\beta, \phi) & 0_{NT \times NT} \\
            0_{N^{2} \times N^{2}} & 0_{NT \times NT} & D^{(3)}(\beta, \phi)
        \end{pmatrix} \, , \label{eq:diagonal_ip_hessian} \\[1em]
    G(\beta, \phi)  =& \, \begin{pmatrix}
            G^{(1, 1)} & G^{(1, 2)}(\beta, \phi) & G^{(1, 3)}(\beta, \phi) \\
            G^{(2, 1)}(\beta, \phi) & G^{(2, 2)} & G^{(2, 3)}(\beta, \phi) \\
            G^{(3, 1)}(\beta, \phi) & G^{(3, 2)}(\beta, \phi) & G^{(3, 3)}
    \end{pmatrix} \, . \nonumber
\end{align}
More specifically,
\begin{align*}
    D^{(1)}(\beta, \phi) =& \, ((w^{(1)})^{\prime} \diag(\mu^{\langle 1 \rangle}(x \beta + w \phi)) w^{(1)}) / \sqrt{NT} \, , \\
    D^{(2)}(\beta, \phi) =& \, ((w^{(2)})^{\prime} \diag(\mu^{\langle 1 \rangle}(x \beta + w \phi)) w^{(2)}) / \sqrt{NT} \, , \\
    D^{(3)}(\beta, \phi) =& \, ((w^{(3)})^{\prime} \diag(\mu^{\langle 1 \rangle}(x \beta + w \phi)) w^{(3)}) / \sqrt{NT} \, , \\
    G^{(1, 1)} =& \, c_{1} (v_{1} v_{1}^{\prime} + v_{2} v_{2}^{\prime}) / \sqrt{NT} \, , \\
    G^{(1, 2)}(\beta, \phi) =& \, ((w^{(1)})^{\prime} \diag(\mu^{\langle 1 \rangle}(x \beta + w \phi)) w^{(2)} + c_{1} v_{1} v_{4}^{\prime}) / \sqrt{NT} \, , \\
    G^{(1, 3)}(\beta, \phi) =& \, ((w^{(1)})^{\prime} \diag(\mu^{\langle 1 \rangle}(x \beta + w \phi)) w^{(3)} + c_{1} v_{2} v_{8}^{\prime}) / \sqrt{NT} \, , \\
    G^{(2, 2)} =& \, c_{1} (v_{4} v_{4}^{\prime} + v_{6} v_{6}^{\prime}) / \sqrt{NT} \, , \\
    G^{(2, 3)}(\beta, \phi) =& \, ((w^{(2)})^{\prime} \diag(\mu^{\langle 1 \rangle}(x \beta + w \phi)) w^{(3)} + c_{1} v_{6} v_{9}^{\prime}) / \sqrt{NT} \, , \\
    G^{(3, 3)} =& \, c_{1} (v_{8} v_{8}^{\prime} + v_{9} v_{9}^{\prime}) / \sqrt{NT} \, .
\end{align*}

\begin{lemma}[Inverse of the incidental parameter Hessian]\label{lemma:inverse_ip_hessian}
    Let Assumption 1 hold. Then, for any $c_{1} > 0$,
    \begin{equation*}
        \norm{(H(\beta, \phi))^{- 1} - (D(\beta, \phi))^{- 1}}_{\max} = \mathcal{O}_{P}((NT)^{- 1 / 2}) \, .
    \end{equation*}
\end{lemma}

\noindent\textbf{Proof of Lemma \ref{lemma:inverse_ip_hessian}.} We adopt the proof strategy used by \textcite{jw2019} and \textcite{wz2021}. Further, we omit the arguments to simplify the notation.

First, since $H$ is non-singular, as implied by Lemma \ref{lemma:ip_hessian}, the following equality holds
\begin{equation}
    \label{eq:ip_hessian_equality}
    H^{- 1} H = H^{- 1} (D + G) = \eye_{\dim(\phi)} \, , 
\end{equation}
where $\dim(\phi) = 2NT + N^2$. Rearranging \eqref{eq:ip_hessian_equality} and exploiting that $H$ is symmetric we get
\begin{align}
    H^{- 1} =& \, D^{- 1} - H^{- 1} G D^{- 1}  \label{eq:ip_hessian_system1} \\
    H^{- 1} =& \, D^{- 1} - D^{- 1} G H^{- 1} \, . \label{eq:ip_hessian_system2}
\end{align}
Substituting \eqref{eq:ip_hessian_system2} into \eqref{eq:ip_hessian_system1}, we get
\begin{equation}
    \label{eq:ip_hessian_expansion}
    H^{- 1} - D^{- 1} = - D^{- 1} G D^{- 1} + D^{- 1} G H^{- 1} G D^{- 1} \, .
\end{equation}

Second, we show that the maximum over the absolute values of all elements in \eqref{eq:ip_hessian_expansion} becomes asymptotically small. By the triangle inequality, the fact that $D$ is a diagonal matrix, and the Cauchy-Schwartz inequality it follows that
\begin{align*}
    \norm{H^{- 1} - D^{- 1}}_{\max} \leq& \, \underset{\{g, h \in \{1, \ldots, 2NT + N^2\}\}}{\max} \big\lvert e_{g}^{\prime} \big(D^{- 1} G D^{- 1}\big) e_{h} \big\rvert + \\
    & \, \underset{\{g, h \in \{1, \ldots, 2NT + N^2\}\}}{\max} \big\lvert e_{g}^{\prime} \big(D^{- 1} G H^{- 1} G D^{- 1}\big) e_{h} \big\rvert \\
    \leq& \, \norm{(\mu^{\langle 1 \rangle})^{- 1}}_{\infty}^{2} \, \underset{\{g, h \in \{1, \ldots, 2NT + N^2\}\}}{\max} \, \big\lvert e_{g}^{\prime} G e_{h} \big\rvert + \\
    & \, \norm{(\mu^{\langle 1 \rangle})^{- 1}}_{\infty}^{2} \, \underset{\{g \in \{1, \ldots, 2NT + N^2\}\}}{\max} \, \norm{G e_{g}}_{2}^{2} \norm{H^{- 1}}_{2} \\
    =& \, \mathcal{O}_{P}((NT)^{- 1 / 2}) \, ,
\end{align*}
where we also used that $\norm{(\mu^{\langle 1 \rangle})^{- 1}}_{\infty} = \mathcal{O}_{P}(1)$, $\max_{g, h} \vert e_{g}^{\prime} G e_{h} \rvert = \mathcal{O}_{P}(1 / \sqrt{NT})$, and $\max_{g} \norm{G e_{g}}_{2}^{2} = \mathcal{O}_{P}(1 / \sqrt{NT})$ are implied by Assumption 1 iii) and the definition of $v$, and that $\norm{H^{- 1}}_{2} = \mathcal{O}_{P}(1)$ by Lemma \ref{lemma:ip_hessian}. \hfill\qedsymbol

\subsection{Asymptotic expansions}
\label{app:asymptotic_expansions}

We follow \textcite{fw2016} and expand the scores of the profile objective function as well as the estimator of the incidental parameters, using Legendre transforms of $L(\beta, \phi)$. To use the Legendre transformation, we need $L(\beta, \phi)$ to be strictly convex in $\phi$. Given Lemma \ref{lemma:ip_hessian}, $\partial_{\phi \phi^{\prime}} L(\beta, \phi) = H(\beta, \phi)$ is positive definite which implies strict convexity in $\phi$. Further, $v^{\prime} \phi^{0} = 0$ by Assumption 1 ii).\vspace{1em}

\noindent\textbf{Legendre transform.} Let $\dim(\phi) = 2NT + N^2$ such that $L(\beta, \phi)\colon \mathcal{D} \subseteq \mathbb{R}^{\dim(\phi)} \mapsto \mathbb{R}$ is a strictly convex function. Then,
\begin{align}
    \widetilde{L}(\beta, \varphi) =& \, \max_{\phi \in \mathcal{D}} (\varphi^{\prime} \phi - L(\beta, \phi)) \, , \label{eq:legendre_transform} \\
    \phi^{\ast}(\beta, \varphi) =& \, \underset{\phi \in \mathcal{D}}{\argmax} (\varphi^{\prime} \phi - L(\beta, \phi)) \, , \label{eq:legendre_transform_argmax}
\end{align}
where 
\begin{equation*}
    \varphi \in \widetilde{\mathcal{D}} = \big\{\varphi \in \mathbb{R}^{\dim(\phi)} \colon \max_{\phi \in \mathcal{D}} (\varphi^{\prime} \phi - L(\beta, \phi)) < \infty\big\} \, .
\end{equation*}
We obtain the profile objective function by evaluating \eqref{eq:legendre_transform} at $\varphi = 0_{\dim(\phi)}$,
\begin{equation*}
    \widetilde{L}(\beta, 0_{\dim(\phi)}) = \min_{\phi \in \mathcal{D}} L(\beta, \phi) = L(\beta, \hat{\phi}(\beta)) \, ,
\end{equation*}
where
\begin{equation*}
    \hat{\phi}(\beta) = \underset{\phi \in \mathcal{D}}{\argmin} L(\beta, \phi) \, .
\end{equation*}
Further, from \eqref{eq:legendre_transform} and \eqref{eq:legendre_transform_argmax}, it follows that 
\begin{equation}
    \label{eq:legendre_transform2}
    \widetilde{L}(\beta, \varphi) = \varphi^{\prime} \phi^{\ast}(\beta, \varphi) - L(\beta, \phi^{\ast}(\beta, \varphi)) \, .
\end{equation}
Furthermore, \eqref{eq:legendre_transform} implies
\begin{equation*}
    \frac{\partial (\varphi^{\prime} \phi - L(\beta, \phi))}{\partial \phi} \biggr\rvert_{\phi = \phi^{\ast}(\beta, \varphi)} = \varphi - \frac{\partial L(\beta, \phi^{\ast}(\beta, \varphi)))}{\partial \phi} = 0 \, .
\end{equation*}
Differentiating both sides with respect to $\beta$ and $\varphi$, we get
\begin{align}
    \frac{\partial}{\partial \beta^{\prime}} \Big(\varphi - \frac{\partial L(\beta, \phi^{\ast}(\beta, \varphi)))}{\partial \phi}\Big) =& \, 0 \, , \nonumber \\
    \frac{\partial}{\partial \varphi^{\prime}} \Big(\varphi - \frac{\partial L(\beta, \phi^{\ast}(\beta, \varphi)))}{\partial \phi}\Big) =& \, 0 \, . \nonumber
\end{align}
Rearranging gives
\begin{align}
    \frac{\partial \phi^{\ast}(\beta, \varphi)}{\partial \beta^{\prime}} =& \, - (H(\beta, \varphi))^{- 1} J(\beta, \varphi) \, , \\
    \frac{\partial \phi^{\ast}(\beta, \varphi)}{\partial \varphi^{\prime}} =& \, (H(\beta, \varphi))^{- 1} \, ,
\end{align}
where 
\begin{align}
    H(\beta, \varphi) =& \, \frac{\partial^{2} L(\beta, \phi^{\ast}(\beta, \varphi))}{\partial \phi \partial \phi^{\prime}} \, , \nonumber \\
    J(\beta, \varphi) =& \, \frac{\partial^{2} L(\beta, \phi^{\ast}(\beta, \varphi))}{\partial \phi \partial \beta^{\prime}} \, . \nonumber
\end{align}
\vspace{1em}

\noindent\textbf{Derivatives of Legendre transform.} We obtain the derivatives for Taylor expansions by differentiating \eqref{eq:legendre_transform2}, 
\begin{align}
    \frac{\partial \widetilde{L}(\beta, \varphi)}{\partial \beta_{k}} =& \, - \, \frac{\partial L(\beta, \phi^{\ast}(\beta, \varphi))}{\partial \beta_{k}} \, , \label{eq:derivatives_legendre} \\
    \frac{\partial \widetilde{L}(\beta, \varphi)}{\partial \varphi} =& \, \phi^{\ast}(\beta, \varphi) \, , \nonumber \\
    \frac{\partial^{2} \widetilde{L}(\beta, \varphi)}{\partial \beta_{k} \partial \beta^{\prime}} =& \, - \, \frac{\partial^{2} L(\beta, \phi^{\ast}(\beta, \varphi))}{\partial \beta_{k} \partial \beta^{\prime}} + (\Phi(\beta, \varphi) e_{k})^{\prime} J(\beta, \varphi) \, , \nonumber \\
    \frac{\partial^{2} \widetilde{L}(\beta, \varphi)}{\partial \beta_{k} \partial \varphi^{\prime}} =& \, - (\Phi(\beta, \varphi) e_{k})^{\prime} \, , \nonumber \\
    \frac{\partial^{2} \widetilde{L}(\beta, \varphi)}{\partial \varphi \partial \varphi^{\prime}} =& \, (H(\beta, \varphi))^{- 1} \, , \nonumber \\
    \frac{\partial^{3} \widetilde{L}(\beta, \varphi)}{\partial \beta_{k} \partial \beta \partial \beta^{\prime}} =& \, - A_{k}(\beta, \varphi) + B_{k}(\beta, \varphi) \Phi(\beta, \varphi) + (B_{k}(\beta, \varphi) \Phi(\beta, \varphi))^{\prime} -  \nonumber \\
    & \, (\Phi(\beta, \varphi))^{\prime} C_{k}(\beta, \varphi) \Phi(\beta, \varphi) \, , \nonumber \\
    \frac{\partial^{3} \widetilde{L}(\beta, \varphi)}{\partial \beta_{k} \partial \beta \partial \varphi^{\prime}} =& \, - B_{k}(\beta, \varphi) (H(\beta, \varphi))^{- 1} + (\Phi(\beta, \varphi))^{\prime}  C_{k}(\beta, \varphi) (H(\beta, \varphi))^{- 1} \, , \nonumber \\ 
    \frac{\partial^{3} \widetilde{L}(\beta, \varphi)}{\partial \beta_{k} \partial \varphi \partial \varphi^{\prime}} =& \, - (H(\beta, \varphi))^{- 1} C_{k}(\beta, \varphi) (H(\beta, \varphi))^{- 1} \, , \nonumber \\
    \frac{\partial^{4} \widetilde{L}(\beta, \varphi)}{\partial \beta_{k} \partial \varphi_{g} \partial \varphi \partial \varphi^{\prime}} =& \, 2 (H(\beta, \varphi))^{- 1} C_{k}(\beta, \varphi) (H(\beta, \varphi))^{- 1} D_{g}(\beta, \varphi) (H(\beta, \varphi))^{- 1} -  \nonumber \\
    & \, (H(\beta, \varphi))^{- 1} E_{k, g}(\beta, \varphi) (H(\beta, \varphi))^{- 1} \, , \nonumber
\end{align}
where
\begin{align*}
    \Phi(\beta, \varphi) =& \, (H(\beta, \varphi))^{- 1} J(\beta, \varphi) \, ,  \\
    A_{k}(\beta, \varphi) =& \, \frac{\partial^{3} L(\beta, \phi^{\ast}(\beta, \varphi))}{\partial \beta_{k} \partial \beta \partial \beta^{\prime}} - \sum_{g = 1}^{\dim(\phi)} \frac{\partial^{3} L(\beta, \phi^{\ast}(\beta, \varphi))}{\partial \phi_{g} \partial \beta \partial \beta^{\prime}} e_{g}^{\prime} \Phi(\beta, \varphi) e_{k} \, ,  \\
    B_{k}(\beta, \varphi) =& \, \frac{\partial^{3} L(\beta, \phi^{\ast}(\beta, \varphi))}{\partial \beta_{k} \partial \beta \partial \phi^{\prime}} - \sum_{g = 1}^{\dim(\phi)} \frac{\partial^{3} L(\beta, \phi^{\ast}(\beta, \varphi))}{\partial \phi_{g} \partial \beta \partial \phi^{\prime}} e_{g}^{\prime} \Phi(\beta, \varphi) e_{k} \, ,  \\
    C_{k}(\beta, \varphi) =& \, \frac{\partial^{3} L(\beta, \phi^{\ast}(\beta, \varphi))}{\partial \beta_{k} \partial \phi \partial \phi^{\prime}} - \sum_{g = 1}^{\dim(\phi)} \frac{\partial^{3} L(\beta, \phi^{\ast}(\beta, \varphi))}{\partial \phi_{g} \partial \phi \partial \phi^{\prime}} e_{g}^{\prime} \Phi(\beta, \varphi) e_{k} \, ,  \\
    D_{g}(\beta, \varphi) =& \, \sum_{h = 1}^{\dim(\phi)} \frac{\partial^{3} L(\beta, \phi^{\ast}(\beta, \varphi))}{\partial \phi_{h} \partial \phi \partial \phi^{\prime}} e_{h}^{\prime} (H(\beta, \varphi))^{- 1} e_{g} \, ,  \\
    E_{k, g}(\beta, \varphi) =& \, \sum_{h = 1}^{\dim(\phi)} \frac{\partial^{4} L(\beta, \phi^{\ast}(\beta, \varphi))}{\partial \beta_{k} \partial \phi_{h} \partial \phi \partial \phi^{\prime}} e_{h}^{\prime} (H(\beta, \varphi))^{- 1} e_{g} \, -    \\
    & \, \sum_{h = 1}^{\dim(\phi)} \Big( \sum_{m = 1}^{\dim(\phi)} \frac{\partial^{4} L(\beta, \phi^{\ast}(\beta, \varphi))}{\partial \phi_{h} \partial \phi_{m} \partial \phi \partial \phi^{\prime}} e_{m}^{\prime} (H(\beta, \varphi))^{- 1} e_{g} \Big) e_{h}^{\prime} \Phi(\beta, \varphi) e_{k} +  \\
    & \, \sum_{h = 1}^{\dim(\phi)} \frac{\partial^{3} L(\beta, \phi^{\ast}(\beta, \varphi))}{\partial \phi_{h} \partial \phi \partial \phi^{\prime}} e_{h}^{\prime} ( (H(\beta, \varphi))^{- 1} D_{g}(\beta, \varphi) \Phi(\beta, \varphi) ) e_{k} -   \\
    & \, \sum_{h = 1}^{\dim(\phi)} \frac{\partial^{3} L(\beta, \phi^{\ast}(\beta, \varphi))}{\partial \phi_{h} \partial \phi \partial \phi^{\prime}} e_{h}^{\prime} \Big( (H(\beta, \varphi))^{- 1} \\
    & \, \Big\{ \sum_{m = 1}^{\dim(\phi)} \frac{\partial^{3} L(\beta, \phi^{\ast}(\beta, \varphi))}{\partial \phi_{m} \partial \phi \partial \beta^{\prime}} e_{m}^{\prime} (H(\beta, \varphi))^{- 1} e_{g} \Big\} \Big) e_{k} \, .
\end{align*}
\vspace{1em}

\noindent\textbf{Expanding the scores of the profile objective function.} We express
\begin{equation*}
    \frac{\partial L(\beta, \hat{\phi}(\beta))}{\partial \beta} = - \frac{\partial \widetilde{L}(\beta, \varphi)}{\partial \beta}\biggr\rvert_{\substack{\beta \, = \, \beta \\ \varphi \, = \, 0}}
\end{equation*}
as first- and second-order Taylor expansions around $\beta = \beta^{0}$ and $\varphi^{0} \in \{\varphi \in \widetilde{\mathcal{D}} \colon \phi^{\ast}(\beta^{0}, \varphi) = \phi_{0}\}$. Since the function to expand is vector-valued, we follow \textcite{fwct2014} and obtain exact forms of Taylor's Theorem by separate expansions for each $k \in \{1, \ldots, K\}$, 
\begin{align}
    \label{eq:first_order_taylor_expansion_legendre_scores}
    \frac{\partial L(\beta, \hat{\phi}(\beta))}{\partial \beta_{k}} =& \, - \frac{\partial \widetilde{L}(\beta, \varphi)}{\partial \beta_{k}}\biggr\rvert_{\substack{\beta \, = \, \beta^{0} \\ \varphi \, = \, \varphi^{0}}} - \, \frac{\partial^{2} \widetilde{L}(\beta, \varphi)}{\partial \beta_{k} \partial \beta^{\prime}}\biggr\rvert_{\substack{\beta \, = \, \check{\beta} \\ \varphi \, = \, \check{\varphi}}} (\beta - \beta^{0}) + \frac{\partial^{2} \widetilde{L}(\beta, \varphi)}{\partial \beta_{k} \partial \varphi^{\prime}}\biggr\rvert_{\substack{\beta \, = \, \check{\beta} \\ \varphi \, = \, \check{\varphi}}} \varphi^{0}
\end{align}
and
\begin{align}
    \label{eq:second_order_taylor_expansion_legendre_scores}
    \frac{\partial L(\beta, \hat{\phi}(\beta))}{\partial \beta_{k}} =& \, - \frac{\partial \widetilde{L}(\beta, \varphi)}{\partial \beta_{k}}\biggr\rvert_{\substack{\beta \, = \, \beta^{0} \\ \varphi \, = \, \varphi^{0}}} - \frac{\partial^{2} \widetilde{L}(\beta, \varphi)}{\partial \beta_{k} \partial \beta^{\prime}}\biggr\rvert_{\substack{\beta \, = \, \beta^{0} \\ \varphi \, = \, \varphi^{0}}} (\beta - \beta^{0}) + \frac{\partial^{2} \widetilde{L}(\beta, \varphi)}{\partial \beta_{k} \partial \varphi^{\prime}}\biggr\rvert_{\substack{\beta \, = \, \beta^{0} \\ \varphi \, = \, \varphi^{0}}} \varphi^{0} \, - \nonumber \\
    & \, \frac{1}{2} (\beta - \beta^{0})^{\prime} \frac{\partial^{3} \widetilde{L}(\beta, \varphi)}{\partial \beta_{k} \partial \beta \partial \beta^{\prime}}\biggr\rvert_{\substack{\beta \, = \, \check{\beta} \\ \varphi \, = \, \check{\varphi}}} (\beta - \beta^{0}) + \frac{1}{2} (\beta - \beta^{0})^{\prime} \frac{\partial^{3} \widetilde{L}(\beta, \varphi)}{\partial \beta_{k} \partial \beta \partial \varphi^{\prime}}\biggr\rvert_{\substack{\beta \, = \, \check{\beta} \\ \varphi \, = \, \check{\varphi}}} \varphi^{0} - \nonumber \\
    & \, \frac{1}{2} (\varphi^{0})^{\prime} \frac{\partial^{3} \widetilde{L}(\beta, \varphi)}{\partial \beta_{k} \partial \varphi \partial \varphi^{\prime}}\biggr\rvert_{\substack{\beta \, = \, \beta^{0} \\ \varphi \, = \, \varphi^{0}}} \varphi^{0} + \frac{1}{6} \sum_{g = 1}^{\dim(\phi)} (\varphi^{0})^{\prime} \frac{\partial^{4} \widetilde{L}(\beta, \varphi)}{\partial \beta_{k} \partial \varphi_{g} \partial \varphi \partial \varphi^{\prime}}\biggr\rvert_{\substack{\beta \, = \, \check{\beta} \\ \varphi \, = \, \check{\varphi}}} \varphi^{0} \varphi_{g}^{0} \, , 
\end{align}
where $\check{\beta}$ and $\check{\varphi}$ are on the line segment between $\beta^{0}$ and $\beta$, and $\varphi^{0}$ and $0$, respectively, and can be different for each $k$.\vspace{1em}

\noindent\textbf{Expanding the estimator for the incidental parameters.} We express
\begin{equation*}
    \hat{\phi}(\beta) = \frac{\partial \widetilde{L}(\beta, \varphi)}{\partial \varphi}\biggr\rvert_{\substack{\beta \, = \, \beta \\ \varphi \, = \, 0}}
\end{equation*}
as first-order Taylor expansions around $\beta = \beta^{0}$ and $\varphi^{0} \in \{\varphi \in \widetilde{\mathcal{D}} \colon \phi^{\ast}(\beta^{0}, \varphi) = \phi_{0}\}$. Again, since the function to expand is vector-valued, we follow \textcite{fwct2014} and obtain exact forms of Taylor's Theorem by separate expansions for each $g \in \{1, \ldots, 2 N T + N^{2}\}$. Let $e_{g}$ denotes a basis vector with a one in the $g$-th coordinate and zeros everywhere else, then 
\begin{equation}
    \label{eq:first_order_taylor_expansion_legendre_phi}
    e_{g}^{\prime} \hat{\phi}(\beta) =  \frac{\partial \widetilde{L}(\beta, \varphi)}{\partial \varphi_{g}}\biggr\rvert_{\substack{\beta \, = \, \beta^{0} \\ \varphi \, = \, \varphi^{0}}} + \, \frac{\partial^{2} \widetilde{L}(\beta, \varphi)}{\partial \varphi_{g} \partial \beta^{\prime}}\biggr\rvert_{\substack{\beta \, = \, \check{\beta} \\ \varphi \, = \, \check{\varphi}}} (\beta - \beta^{0}) - \frac{\partial^{2} \widetilde{L}(\beta, \varphi)}{\partial \varphi_{g} \partial \varphi^{\prime}}\biggr\rvert_{\substack{\beta \, = \, \check{\beta} \\ \varphi \, = \, \check{\varphi}}} \varphi^{0} \, ,
\end{equation}
where $\check{\beta}$ and $\check{\varphi}$ are on the line segment between $\beta^{0}$ and $\beta$, and $\varphi^{0}$ and $0$, respectively, and can be different for each $g$.\vspace{1em}

\noindent\textbf{Derivatives of the log-likelihood function.} We define the required derivatives of the log-likelihood function, defined in \eqref{eq:objective_function}, before presenting explicit expressions for the first- and second-order expansions based on Legendre transforms.
\begin{align}
    \frac{\partial L(\beta, \phi)}{\partial \beta} =& \, - (x^{\prime} u(\beta, \phi)) / \sqrt{NT} \, , \label{eq:derivatives_objective_function} \\
    \frac{\partial L(\beta, \phi)}{\partial \phi} =& \, - (w^{\prime} u(\beta, \phi) - v v^{\prime} \phi) / \sqrt{NT} \, , \nonumber \\
    \frac{\partial^{2} L(\beta, \phi)}{\partial \beta \partial \beta^{\prime}} =& \, (x^{\prime} \diag(\mu^{\langle 1 \rangle}(x \beta + w \phi)) x) / \sqrt{NT} \, , \nonumber  \\
    \frac{\partial^{2} L(\beta, \phi)}{\partial \beta \partial \phi^{\prime}} =& \, (x^{\prime} \diag(\mu^{\langle 1 \rangle}(x \beta + w \phi)) w) / \sqrt{NT} \, , \nonumber \\
    \frac{\partial^{2} L(\beta, \phi)}{\partial \phi \partial \phi^{\prime}} =& \, (w^{\prime} \diag(\mu^{\langle 1 \rangle}(x \beta + w \phi)) w + v v^{\prime}) / \sqrt{NT} \, , \nonumber \\
    \frac{\partial^{3} L(\beta, \phi)}{\partial \beta_{k} \partial \beta \partial \beta^{\prime}} =& \, (x^{\prime} \diag(\mu^{\langle 2 \rangle}(x \beta + w \phi) \odot x  e_{k}) x) / \sqrt{NT} \, , \nonumber \\
    \frac{\partial^{3} L(\beta, \phi)}{\partial \beta_{k} \partial \beta \partial \phi^{\prime}} =& \, (x^{\prime} \diag(\mu^{\langle 2 \rangle}(x \beta + w \phi) \odot x e_{k}) w) / \sqrt{NT} \, , \nonumber \\
    \frac{\partial^{3} L(\beta, \phi)}{\partial \beta_{k} \partial \phi \partial \phi^{\prime}} =& \, (w^{\prime} \diag(\mu^{\langle 2 \rangle}(x \beta + w \phi) \odot x  e_{k}) w) / \sqrt{NT} \, , \nonumber \\
    \frac{\partial^{3} L(\beta, \phi)}{\partial \phi_{g} \partial \beta \partial \beta^{\prime}} =& \, (x^{\prime} \diag(\mu^{\langle 2 \rangle}(x \beta + w \phi) \odot w e_{g}) x) / \sqrt{NT} \, , \nonumber \\
    \frac{\partial^{3} L(\beta, \phi)}{\partial \phi_{g} \partial \beta \partial \phi^{\prime}} =& \, (x^{\prime} \diag(\mu^{\langle 2 \rangle}(x \beta + w \phi) \odot w e_{g}) w) / \sqrt{NT} \, , \nonumber \\
    \frac{\partial^{3} L(\beta, \phi)}{\partial \phi_{h} \partial \phi \partial \phi^{\prime}} =& \, (w^{\prime} \diag(\mu^{\langle 2 \rangle}(x \beta + w \phi) \odot w e_{g}) w) / \sqrt{NT} \, , \nonumber \\
    \frac{\partial^{4} L(\beta, \phi)}{\partial \beta_{k} \partial \beta_{l} \partial \beta \partial \beta^{\prime}} =& \, (x^{\prime} \diag(\mu^{\langle 3 \rangle}(x \beta + w \phi) \odot x e_{k} \odot x e_{l}) x) / \sqrt{NT} \, , \nonumber \\
    \frac{\partial^{4} L(\beta, \phi)}{\partial \beta_{k} \partial \beta_{l} \partial \beta \partial \phi^{\prime}} =& \, (x^{\prime} \diag(\mu^{\langle 3 \rangle}(x \beta + w \phi) \odot x e_{k} \odot x e_{l}) w) / \sqrt{NT} \, , \nonumber \\
    \frac{\partial^{4} L(\beta, \phi)}{\partial \beta_{k} \partial \beta_{l} \partial \phi \partial \phi^{\prime}} =& \, (w^{\prime} \diag(\mu^{\langle 3 \rangle}(x \beta + w \phi) \odot x e_{k} \odot x e_{l}) w) / \sqrt{NT} \, , \nonumber \\
    \frac{\partial^{4} L(\beta, \phi)}{\partial \beta_{k} \partial \phi_{g} \partial \beta \partial \beta^{\prime}} =& \, (x^{\prime} \diag(\mu^{\langle 3 \rangle}(x \beta + w \phi) \odot x e_{k} \odot w e_{g}) x) / \sqrt{NT} \, , \nonumber \\
    \frac{\partial^{4} L(\beta, \phi)}{\partial \beta_{k} \partial \phi_{g} \partial \beta \partial \phi^{\prime}} =& \, (x^{\prime} \diag(\mu^{\langle 3 \rangle}(x \beta + w \phi) \odot x e_{k} \odot w e_{g}) w) / \sqrt{NT} \, , \nonumber \\
    \frac{\partial^{4} L(\beta, \phi)}{\partial \beta_{k} \partial \phi_{g} \partial \phi \partial \phi^{\prime}} =& \, (w^{\prime} \diag(\mu^{\langle 3 \rangle}(x \beta + w \phi) \odot x e_{k} \odot w e_{g}) w) / \sqrt{NT} \, , \nonumber \\
    \frac{\partial^{4} L(\beta, \phi)}{\partial \phi_{g} \partial \phi_{h} \partial \beta \partial \beta^{\prime}} =& \, (x^{\prime} \diag(\mu^{\langle 3 \rangle}(x \beta + w \phi) \odot w e_{g} \odot w e_{h}) x) / \sqrt{NT} \, , \nonumber \\
    \frac{\partial^{4} L(\beta, \phi)}{\partial \phi_{g} \partial \phi_{h} \partial \beta \partial \phi^{\prime}} =& \, (x^{\prime} \diag(\mu^{\langle 3 \rangle}(x \beta + w \phi) \odot w e_{g} \odot w e_{h}) w) / \sqrt{NT} \, , \nonumber \\
    \frac{\partial^{4} L(\beta, \phi)}{\partial \phi_{g} \partial \phi_{h} \partial \phi \partial \phi^{\prime}} =& \, (w^{\prime} \diag(\mu^{\langle 3 \rangle}(x \beta + w \phi) \odot w e_{g} \odot w e_{h}) w) / \sqrt{NT} \, . \nonumber 
\end{align}
\vspace{1em}

\noindent\textbf{Explicit expressions for the first-order expansions.} We use the derivatives of the Legendre transform given in \eqref{eq:derivatives_legendre} and the explicit expressions for the derivatives of the log-likelihood function given in \eqref{eq:derivatives_objective_function}. To make the first-order expansions more concise, we stack the corresponding equations, \eqref{eq:first_order_taylor_expansion_legendre_scores} and \eqref{eq:first_order_taylor_expansion_legendre_phi}.

\begin{lemma}[First-order expansions for $\hat{\beta}$ and $\hat{\phi}(\hat{\beta})$]\label{lemma:first_order_expansions}
    Let Assumption 1 hold. Then, i)
    \begin{equation*}
        \frac{1}{\sqrt{NT}} (\check{M} x)^{\prime} \diag(\check{\mu^{\langle 1 \rangle}}) x (\hat{\beta} - \beta^{0}) = \frac{1}{\sqrt{NT}} (\check{M} x)^{\prime} u 
    \end{equation*}
    and ii)
    \begin{equation*}
        \hat{\phi}(\hat{\beta}) - \phi^{0} =  - \frac{1}{\sqrt{NT}} \check{H}^{- 1} w^{\prime} \diag(\check{\mu}^{\langle 1 \rangle}) x (\hat{\beta} - \beta^{0}) + \frac{1}{\sqrt{NT}} (\check{H})^{- 1} w^{\prime} u \, .
    \end{equation*}
\end{lemma}

\noindent\textbf{Proof of Lemma \ref{lemma:first_order_expansions}.} The Lemma follows immediately by plugin in the expressions for the derivatives \eqref{eq:derivatives_objective_function} into \eqref{eq:first_order_taylor_expansion_legendre_scores} and \eqref{eq:first_order_taylor_expansion_legendre_phi}, the fact that the definition of $\hat{\beta}$ implies that $\partial_{\beta} L(\hat{\beta}, \hat{\phi}(\hat{\beta})) = 0$, and some rearrangement.\hfill\qedsymbol
\vspace{1em}

\noindent\textbf{Explicit expression for the second-order expansion.} Again, we use the derivatives of the Legendre transform given in \eqref{eq:derivatives_legendre} and the explicit expressions for the derivatives of the log-likelihood function given in \eqref{eq:derivatives_objective_function}.

\begin{lemma}[Second-order expansion for $\hat{\beta}$]\label{lemma:second_order_expansion}
    Let Assumption 1 hold. Then,
    \begin{equation*}
        N\sqrt{T} (\hat{\beta} - \beta^{0}) = W^{- 1} U^{(0)} + W^{- 1} U^{(1)} + o_{p}(1) \, ,
    \end{equation*}
    where 
    \begin{align*}
        U^{(0)} =& \, \frac{1}{N \sqrt{T}} (M x)^{\prime} u \overset{d}{\rightarrow} \mathcal{N}(0, \overline{W}) \, ,\\
        U^{(1)} =& \, - \frac{1}{2 N \sqrt{T}} (\mathbb{Q} u)^{\prime} \diag(\mu^{\langle 2 \rangle} \odot \mathbb{Q} u)  M x \, ,
    \end{align*}
    and $\overline{W} = \EX{W}$.
\end{lemma}

\noindent\textbf{Proof of Lemma \ref{lemma:second_order_expansion}.} We plug the expressions for the derivatives \eqref{eq:derivatives_objective_function} into \eqref{eq:second_order_taylor_expansion_legendre_scores} and use the fact that the definition of $\hat{\beta}$ implies that $\partial_{\beta} L(\hat{\beta}, \hat{\phi}(\hat{\beta})) = 0$. Then, for each $k \in \{1, \ldots, K\}$, we have
\begin{align}
    0 =& \, - \frac{1}{\sqrt{NT}} (M x e_{k})^{\prime} u + \frac{1}{\sqrt{NT}} (M x e_{k})^{\prime} \diag(\mu^{\langle 1 \rangle}) x (\hat{\beta} - \beta^{0}) + \label{eq:score_expansion_second_order}\\
    & \, \frac{1}{2 \sqrt{NT}} (\mathbb{Q} u)^{\prime} \diag(\mu^{\langle 2 \rangle} \odot M x e_{k}) \mathbb{Q} u + \sqrt{N} \check{R}_{k}(\hat{\beta}) \, , \nonumber
\end{align}
where
\begin{align*}
    \check{R}_{k}(\hat{\beta}) =& \, \frac{1}{2 N\sqrt{T}} (\hat{\beta} - \beta^{0})^{\prime} (\check{M} x)^{\prime} \diag(\check{\mu}^{\langle 2 \rangle} \odot \check{M} x e_{k}) \check{M} x (\hat{\beta} - \beta^{0}) + \\
    & \, \frac{1}{2 N\sqrt{T}} (\hat{\beta} - \beta^{0})^{\prime} (\check{M} x)^{\prime} \diag(\check{\mu}^{\langle 2 \rangle} \odot \check{M} x e_{k}) \check{\mathbb{Q}} u + \\
    & \, \frac{1}{3 N\sqrt{T}} (\check{\mathbb{Q}} u)^{\prime} \diag(\check{\mu}^{\langle 2 \rangle} \odot \check{M} x e_{k}) \check{\mathbb{Q}} \diag(\check{\mu}^{\langle 2 \rangle} \odot \check{\mathbb{Q}} u) \check{\mathbb{Q}} u + \\
    & \, \frac{1}{6 N\sqrt{T}} (\check{\mathbb{Q}} u)^{\prime} \diag(\check{\mu}^{\langle 3 \rangle} \odot \check{M} x e_{k} \odot \check{\mathbb{Q}} u) \check{\mathbb{Q}} u - \\
    & \, \frac{1}{6 N\sqrt{T}} (\check{\mathbb{Q}} u)^{\prime} \diag(\check{\mu}^{\langle 2 \rangle} \odot \check{\mathbb{Q}} \diag(\check{\mu}^{\langle 2 \rangle} \odot \check{\mathbb{Q}} u) \check{M} x e_{k} ) \check{\mathbb{Q}} u \\
    =& \, \check{R}_{k}^{(1)}(\hat{\beta}) + \check{R}_{k}^{(2)}(\hat{\beta}) + \check{R}_{k}^{(3)} + \check{R}_{k}^{(4)} + \check{R}_{k}^{(5)}
\end{align*}
is the $k$-th element of a $K$-dimensional vector $\check{R}(\hat{\beta})$. Using Lemma \ref{lemma:profile_hessian} and re-arranging \eqref{eq:score_expansion_second_order} yields
\begin{equation*}
    N\sqrt{T} (\hat{\beta} - \beta^{0}) =  W^{- 1} U^{(0)} + W^{- 1} U^{(1)} + W^{- 1} \check{R}(\hat{\beta}) \, ,
\end{equation*}
where
\begin{align*}
    W =& \, \frac{1}{N^2T} (M x)^{\prime} \diag(\mu^{\langle 1 \rangle}) x \, ,\\
    U^{(0)} =& \, \frac{1}{N \sqrt{T}} (M x)^{\prime} u \, ,\\
    U^{(1)} =& \, - \frac{1}{2 N \sqrt{T}} (\mathbb{Q} u)^{\prime} \diag(\mu^{\langle 2 \rangle} \odot \mathbb{Q} u)  M x \, .
\end{align*}

From Lemma \ref{lemma:clt}, it follows that $U^{(0)} \overset{d}{\rightarrow} \mathcal{N}(0, \overline{W})$, where $\overline{W} = \EX{W}$. Further, we use the properties of vector norms and the triangle inequality to decompose the remainder term,
\begin{align}
    \norm{\check{R}(\hat{\beta})}_{2} \leq& \, \norm{\check{R}(\hat{\beta})}_{1} \label{eq:decomposition_remainder_term} \\
    \leq& \, \sum_{k = 1}^{K} \big\lvert\check{R}_{k}^{(1)}(\hat{\beta})\big\rvert + \sum_{k = 1}^{K} \big\lvert\check{R}_{k}^{(2)}(\hat{\beta})\big\rvert + \sum_{k = 1}^{K} \big\lvert\check{R}_{k}^{(3)}\big\rvert + \sum_{k = 1}^{K} \big\lvert\check{R}_{k}^{(4)}\big\rvert + \sum_{k = 1}^{K} \big\lvert\check{R}_{k}^{(5)}\big\rvert \nonumber \, .
\end{align}
In the following, we use Lemmas \ref{lemma:linear_operator}, \ref{lemma:bound_second_order} ii), \ref{lemma:profile_hessian}, and \ref{lemma:convergence_rates} i), along with Assumption 1 iii), to bound the five components in \eqref{eq:decomposition_remainder_term}. For the first component, we have
\begin{align*}
     \big\lvert \check{R}^{(1)}_{k}(\hat{\beta}) \big\rvert =& \, \frac{1}{2 N \sqrt{T}} \big\lvert (\hat{\beta} - \beta^{0})^{\prime} ((\check{M} x)^{\prime} \diag(\check{\mu}^{\langle 2 \rangle} \odot \check{M} x e_{k}) \check{M} x) (\hat{\beta} - \beta^{0}) \big\rvert \\
     \leq& \, \frac{1}{2 N \sqrt{T}} \norm{\hat{\beta} - \beta^{0}}_{2}^{2} \norm{(\check{M} x)^{\prime} \diag(\check{\mu}^{\langle 2 \rangle} \odot \check{M} x e_{k}) \check{M} x}_{2} \\
     \leq& \,\sqrt{K}  N \sqrt{T} \norm{\hat{\beta} - \beta^{0}}_{2}^{2}\norm{\check{\mu}^{\langle 2 \rangle}}_{\infty} \norm{\check{M} x}_{\max}^{3} \\
     =& \, o_{P}(1) \, ,
\end{align*}
where used the quadratic inequality along with the properties of vector norms, $\norm{\hat{\beta} - \beta^{0}}_{2}^{2} = \mathcal{O}_{P}(1 / N^2 T)$, $\norm{\check{\mu}^{\langle 2 \rangle}}_{\infty} = \mathcal{O}_{P}(1)$, and $\norm{\check{M} x}_{\max} = \mathcal{O}_{P}(1)$. For the second component, we have
\begin{align*}
    \big\lvert \check{R}^{(2)}_{k}(\hat{\beta}) \big\rvert =& \, \frac{1}{N \sqrt{T}} \big\lvert (\hat{\beta} - \beta^{0})^{\prime} (\check{M} x)^{\prime} \diag(\check{\mu}^{\langle 2 \rangle} \odot \check{M} x e_{k}) \check{\mathbb{Q}} u \big\rvert \\
    \leq& \,\sqrt{K}  N \sqrt{T} \norm{\hat{\beta} - \beta^{0}}_{2} \norm{\check{\mathbb{Q}} u}_{\infty} \norm{\check{\mu}^{\langle 2 \rangle}}_{\infty} \norm{\check{M} x}_{\max}^{2} \\
    =& \, o_{P}(1) \, ,
\end{align*}
where used Hoelder's inequality along with the properties of vector norms, $\norm{\hat{\beta} - \beta^{0}}_{2} = \mathcal{O}_{P}(1 / N \sqrt{T})$, $\norm{\check{\mathbb{Q}} u}_{\infty} = \mathcal{O}_{P}((N T)^{- 1/ 4})$, $\norm{\check{\mu}^{\langle 2 \rangle}}_{\infty} = \mathcal{O}_{P}(1)$, and $\norm{\check{M} x}_{\max} = \mathcal{O}_{P}(1)$. For the third component, we have
\begin{align*}
    \big\lvert \check{R}^{(3)}_{k} \big\rvert =& \, \frac{1}{3 N\sqrt{T}} \big\lvert(\check{\mathbb{Q}} u)^{\prime} \diag(\check{\mu}^{\langle 2 \rangle} \odot \check{M} x e_{k}) \check{\mathbb{Q}} \diag(\check{\mu}^{\langle 2 \rangle} \odot \check{\mathbb{Q}} u) \check{\mathbb{Q}} u \big\rvert \\
    \leq& \, \frac{1}{N \sqrt{T}} \big\lvert (\check{\mathbb{Q}} u)^{\prime} \check{\mathbb{Q}} \diag(\check{\mathbb{Q}} u) \check{\mathbb{Q}} u \big\rvert \norm{\check{\mu}^{\langle 2 \rangle}}_{\infty}^{2} \norm{\check{M} x}_{\max} \\
    \leq& \, \frac{1}{N \sqrt{T}} \big\lvert (\check{\mathbb{Q}} u)^{\prime} \diag(\check{\mathbb{Q}} u) \check{\mathbb{Q}} u \big\rvert \norm{(\check{\mu}^{\langle 1 \rangle})^{- 1}}_{\infty} \norm{\check{\mu}^{\langle 2 \rangle}}_{\infty}^{2} \norm{\check{M} x}_{\max} \\
    =& \, o_{P}(1) \, ,
\end{align*}
where we used that $\check{\mathbb{Q}} \diag(\check{\mu}^{\langle 1 \rangle}) \check{\mathbb{Q}} = \check{\mathbb{Q}}$, $(\check{\mathbb{Q}} u)^{\prime} \diag(\check{\mathbb{Q}} u) \check{\mathbb{Q}} u = \mathcal{O}_{P}(\sqrt{NT})$, $\norm{(\check{\mu}^{\langle 1 \rangle})^{- 1}}_{\infty} = \mathcal{O}_{P}(1)$, $\norm{\check{\mu}^{\langle 2 \rangle}}_{\infty} = \mathcal{O}_{P}(1)$, and $\norm{\check{M} x}_{\max} = \mathcal{O}_{P}(1)$. For the fourth component, we have
\begin{align*}
    \big\lvert \check{R}^{(4)}_{k} \big\rvert =& \, \frac{1}{6 N\sqrt{T}} \big\lvert (\check{\mathbb{Q}} u)^{\prime} \diag(\check{\mu}^{\langle 3 \rangle} \odot \check{M} x e_{k} \odot \check{\mathbb{Q}} u) \check{\mathbb{Q}} u \big\rvert \\
    \leq& \, \frac{1}{N \sqrt{T}} \big\lvert (\check{\mathbb{Q}} u)^{\prime} \diag(\check{\mathbb{Q}} u) \check{\mathbb{Q}} u \big\rvert \norm{\check{\mu}^{\langle 3 \rangle}}_{\infty} \norm{\check{M} x}_{\max} \\
    =& \, o_{P}(1) \, ,
\end{align*}
where we used that $(\check{\mathbb{Q}} u)^{\prime} \diag(\check{\mathbb{Q}} u) \check{\mathbb{Q}} u = \mathcal{O}_{P}(\sqrt{NT})$, $\norm{\check{\mu}^{\langle 3 \rangle}}_{\infty} = \mathcal{O}_{P}(1)$, and $\norm{\check{M} x}_{\max} = \mathcal{O}_{P}(1)$. For the fifth component, we have
\begin{align*}
    \big\lvert \check{R}^{(5)}_{k} \big\rvert =& \, \frac{1}{6 N \sqrt{T}} \big\lvert (\check{\mathbb{Q}} u)^{\prime} \diag(\check{\mu}^{\langle 2 \rangle} \odot \check{\mathbb{Q}} \diag(\check{\mu}^{\langle 2 \rangle} \odot \check{\mathbb{Q}} u) \check{M} x e_{k} ) \check{\mathbb{Q}} u \big\rvert \\
    \leq& \, \frac{1}{N \sqrt{T}} \big\lvert (\check{\mathbb{Q}} u)^{\prime} \diag(\check{\mathbb{Q}} u) \check{\mathbb{Q}} u \big\rvert \norm{(\check{\mu}^{\langle 1 \rangle})^{- 1}}_{\infty} \norm{\check{\mu}^{\langle 2 \rangle}}_{\infty}^{2} \norm{\check{M} x}_{\max} \\
    =& \, o_{P}(1) \, ,
\end{align*}
where we used that $\check{\mathbb{Q}} \diag(\check{\mu}^{\langle 1 \rangle}) \check{\mathbb{Q}} = \check{\mathbb{Q}}$, $(\check{\mathbb{Q}} u)^{\prime} \diag(\check{\mathbb{Q}} u) \check{\mathbb{Q}} u = \mathcal{O}_{P}(\sqrt{NT})$, $\norm{(\check{\mu}^{\langle 1 \rangle})^{- 1}}_{\infty} = \mathcal{O}_{P}(1)$, $\norm{\check{\mu}^{\langle 2 \rangle}}_{\infty} = \mathcal{O}_{P}(1)$, and $\norm{\check{M} x}_{\max} = \mathcal{O}_{P}(1)$. Bringing all components together and using that $\norm{W^{- 1}}_{2} = \mathcal{O}_{P}(1)$, we conclude that $W^{- 1} \check{R}(\hat{\beta}) = o_{P}(1)$. \hfill\qedsymbol

\subsection{Linear operator}
\label{app:linear_operator}

Let
\begin{equation}
    \label{eq:linear_operator}
    \mathbb{Q}(\beta, \phi) = ((w^{(1)}, w^{(2)}, w^{(3)}) (H(\beta, \phi))^{- 1} (w^{(1)}, w^{(2)}, w^{(3)})^{\prime}) / \sqrt{NT} \, , 
\end{equation}
be the linear operator matrix. Extending \eqref{eq:linear_operator}, we get
\begin{align}
    \mathbb{Q}(\beta, \phi) =& \, (w ((H(\beta, \phi))^{- 1} - (D(\beta, \phi))^{- 1} + (D(\beta, \phi))^{- 1}) w^{\prime}) / \sqrt{NT} \label{eq:decomposition_linear_operator} \\
    =& \, (w ((D(\beta, \phi))^{- 1} w^{\prime}) / \sqrt{NT} + (w ((H(\beta, \phi))^{- 1} - (D(\beta, \phi))^{- 1}) w^{\prime}) / \sqrt{NT} \nonumber \\
    =& \, \mathbb{Q}_{\phi}(\beta, \phi) + \mathbb{Q}_{R}(\beta, \phi) \nonumber \\ 
    =& \, \mathbb{Q}_{\alpha}(\beta, \phi) + \mathbb{Q}_{\gamma}(\beta, \phi) + \mathbb{Q}_{\rho}(\beta, \phi) + \mathbb{Q}_{R}(\beta, \phi) \, , \nonumber
\end{align}
where 
\begin{align*}
    \mathbb{Q}_{\alpha}(\beta, \phi) =& \, w^{(1)} ((w^{(1)})^{\prime} \diag(\mu^{\langle 1 \rangle}(x \beta + w \phi)) w^{(1)})^{- 1} (w^{(1)})^{\prime} \, , \\
    \mathbb{Q}_{\gamma}(\beta, \phi) =& \, w^{(2)} ((w^{(2)})^{\prime} \diag(\mu^{\langle 1 \rangle}(x \beta + w \phi)) w^{(2)})^{- 1} (w^{(2)})^{\prime} \, , \\
    \mathbb{Q}_{\rho}(\beta, \phi) =& \, w^{(3)} ((w^{(3)})^{\prime} \diag(\mu^{\langle 1 \rangle}(x \beta + w \phi)) w^{(3)})^{- 1} (w^{(3)})^{\prime} \, , \\
    \mathbb{Q}_{R}(\beta, \phi) =& \, (w ((H(\beta, \phi))^{- 1} - (D(\beta, \phi))^{- 1}) w^{\prime}) / \sqrt{NT} \, .
\end{align*}
Further, when multiplied by a vector, $\mathbb{Q}_{\alpha}(\beta, \phi)$, $\mathbb{Q}_{\gamma}(\beta, \phi)$, and $\mathbb{Q}_{\rho}(\beta, \phi)$ result in convenient scalar expressions. Let $z \in \mathbb{R}^{N^2T}$ be an arbitrary vector, then
\begin{align*}
    (\mathbb{Q}_{\alpha}(\beta, \phi) z)_{ijt} =& \, \frac{\sum_{j^{\prime} = 1}^{N} z_{ij^{\prime}t}}{\sum_{j^{\prime} = 1}^{N} \mu^{\langle 1 \rangle}(x_{ij^{\prime}t}^{\prime} \beta + w_{ij^{\prime}t}^{\prime} \phi)} \, , \\
    (\mathbb{Q}_{\gamma}(\beta, \phi) z)_{ijt} =& \, \frac{\sum_{i^{\prime} = 1}^{N} z_{i^{\prime}jt}}{\sum_{i^{\prime} = 1}^{N} \mu^{\langle 1 \rangle}(x_{i^{\prime}jt}^{\prime} \beta + w_{i^{\prime}jt}^{\prime} \phi)} \, , \\
    (\mathbb{Q}_{\rho}(\beta, \phi) z)_{ijt} =& \, \frac{\sum_{t^{\prime} = 1}^{T} z_{ijt^{\prime}}}{\sum_{t^{\prime} = 1}^{T} \mu^{\langle 1 \rangle}(x_{ijt^{\prime}}^{\prime} \beta + w_{ijt^{\prime}}^{\prime} \phi)} \, .
\end{align*}

Next, we bound the maximum value of $\mathbb{Q}_{\phi}(\beta, \phi) u$ and $\mathbb{Q}_{R}(\beta, \phi) u$.

\begin{lemma}[Bounds for linear operator]\label{lemma:linear_operator}
    Let Assumption 1 hold. Then,
    \begin{enumerate}[i)]
        \item $\norm{\mathbb{Q}_{\phi}(\beta, \phi) u}_{\infty} = \mathcal{O}_{P}((NT)^{- 1 / 4})$,
        \item $\norm{\mathbb{Q}_{R}(\beta, \phi) u}_{\infty} = \mathcal{O}_{P}((NT)^{- 3 / 4})$.
    \end{enumerate}
\end{lemma}

\noindent\textbf{Proof of Lemma \ref{lemma:linear_operator}.} We bound the maximum over the absolute values of all elements in $\mathbb{Q}_{\phi}(\beta, \phi) u$ and $\mathbb{Q}_{R}(\beta, \phi) u$. We omit the arguments to simplify the notation.

By Hoelder's inequality it follows that,
\begin{align*}
    \norm{\mathbb{Q}_{\phi} u}_{\infty} =& \, \underset{\{n \in \{1, \ldots, N^2T\}\}}{\max} \big\lvert e_{n}^{\prime} w D^{- 1} w^{\prime} u \big\rvert / \sqrt{NT} \\
    \leq& \, \underset{\{n \in \{1, \ldots, N^2T\}\}}{\max} \lvert e_{n}^{\prime} w w^{\prime} u \rvert \, \norm{(\mu^{\langle 1 \rangle})^{- 1}}_{\infty} \, / \sqrt{NT} \\
    \leq& \, \underset{\{n \in \{1, \ldots, N^2T\}\}}{\max} \norm{w^{\prime} e_{n}}_{1} \norm{w^{\prime} u}_{\infty} \norm{(\mu^{\langle 1 \rangle})^{- 1}}_{\infty} \, / \sqrt{NT} \\
    =& \, \norm{w}_{\infty} \norm{w^{\prime} u}_{\infty} \norm{(\mu^{\langle 1 \rangle})^{- 1}}_{\infty} \, / \sqrt{NT} \\
    =& \, \mathcal{O}_{P}((NT)^{- 1 / 4}) \, ,
\end{align*}
where $\norm{w}_{\infty} = 3$, $\norm{w^{\prime} u}_{\infty} = \mathcal{O}_{P}((NT)^{1 / 4})$ by Lemma \ref{lemma:ip_score}, and $\norm{(\mu^{\langle 1 \rangle})^{- 1}}_{\infty} = \mathcal{O}_{P}(1)$. Further,
\begin{align*}
    \norm{\mathbb{Q}_{R} u}_{\infty} =& \, \underset{\{n \in \{1, \ldots, N^2T\}\}}{\max} \big\lvert e_{n}^{\prime} w (H^{- 1} - D^{- 1}) w^{\prime} u \big\rvert / \sqrt{NT} \\
    \leq& \, \norm{w}_{\infty} \norm{w^{\prime} u}_{\infty} \, \norm{H^{- 1} - D^{- 1}}_{\max} / \sqrt{NT} \\
    =& \, \mathcal{O}_{P}((NT)^{- 3 / 4}) \, ,
\end{align*}
where $\norm{H^{- 1} - D^{- 1}}_{\max} = \mathcal{O}_{P}(1 / \sqrt{NT})$ by Lemma \ref{lemma:inverse_ip_hessian}. \hfill\qedsymbol\vspace{1em}

Finally, we provide bounds that depend on the linear operator and are useful for the second-order expansion in Lemma \ref{lemma:second_order_expansion}.

\begin{lemma}[Bounds for second-order expansion]\label{lemma:bound_second_order}
    Let Assumption 1 hold. Then,
    \begin{enumerate}[i)]
        \item $(\mathbb{Q}(\beta, \phi) u)^{\prime} \mathbb{Q}(\beta, \phi) u = \mathcal{O}_{P}(NT)$,
        \item $(\mathbb{Q}(\beta, \phi) u)^{\prime} \diag(\mathbb{Q}(\beta, \phi) u) \mathbb{Q}(\beta, \phi) u = \mathcal{O}_{P}(\sqrt{NT})$.
    \end{enumerate}
\end{lemma}

\noindent\textbf{Proof of Lemma \ref{lemma:bound_second_order}.} As in the proof of the previous Lemma, we omit the arguments to simplify the notation. For i), using Loeve's inequality, we have
\begin{align*}
    \big\lvert (\mathbb{Q} u)^{\prime} \mathbb{Q} u \big\rvert =& \, \Big\lvert \sum_{i = 1}^{N} \sum_{j = 1}^{N} \sum_{t = 1}^{T} ((\mathbb{Q} u)_{ijt})^{2} \Big\rvert \\
    \leq& \, 2 \Big\lvert  \sum_{i = 1}^{N} \sum_{j = 1}^{N} \sum_{t = 1}^{T} ((\mathbb{Q}_{\phi} u)_{ijt})^{2} \Big\rvert + 2 \Big\lvert \sum_{i = 1}^{N} \sum_{j = 1}^{N} \sum_{t = 1}^{T} ((\mathbb{Q}_{R} u)_{ijt})^{2} \Big\rvert\\
    \leq& \, 2 \big\lvert (\mathbb{Q}_{\phi} u)^{\prime} \mathbb{Q}_{\phi} u \big\rvert + 2 N^2 T \, \norm{\mathbb{Q}_{R} u}_{\infty}^{2} \, .
\end{align*}
Further, since $\mathbb{Q}_{\phi} = \mathbb{Q}_{\alpha} + \mathbb{Q}_{\gamma} + \mathbb{Q}_{\rho}$, using again Loeve's inequality, it follows that
\begin{align*}
    \big\lvert (\mathbb{Q}_{\phi} u)^{\prime} \mathbb{Q}_{\phi} u \big\rvert \leq& \, 3 \big\lvert (\mathbb{Q}_{\alpha} u)^{\prime} \mathbb{Q}_{\alpha} u \big\rvert + 3 \big\lvert (\mathbb{Q}_{\gamma} u)^{\prime} \mathbb{Q}_{\gamma} u \big\rvert + 3 \big\lvert (\mathbb{Q}_{\rho} u)^{\prime} \mathbb{Q}_{\rho} u \big\rvert \\
    =& \, 3 N \bigg\lvert  \sum_{i = 1}^{N} \sum_{t = 1}^{T} \frac{(\sum_{j = 1}^{N} u_{ijt})^2}{(\sum_{j = 1}^{N} \mu^{\langle 1 \rangle}_{ijt})^2} \bigg\rvert + 3 N \bigg\lvert  \sum_{j = 1}^{N} \sum_{t = 1}^{T} \frac{(\sum_{i = 1}^{N} u_{ijt})^2}{(\sum_{i = 1}^{N} \mu^{\langle 1 \rangle}_{ijt})^2} \bigg\rvert + \\
    & \, 3 T \bigg\lvert  \sum_{i = 1}^{N} \sum_{j = 1}^{N} \frac{(\sum_{t = 1}^{T} u_{ijt})^2}{(\sum_{t = 1}^{T} \mu^{\langle 1 \rangle}_{ijt})^2} \bigg\rvert \\
    \leq& \, \frac{3}{N} \, \norm{(\mu^{\langle 1 \rangle})^{- 1}}_{\infty} \, \bigg\lvert  \sum_{i = 1}^{N} \sum_{t = 1}^{T} \big(\sum_{j = 1}^{N} u_{ijt}\big)^2 \bigg\rvert + \\
    & \, \frac{3}{N} \, \norm{(\mu^{\langle 1 \rangle})^{- 1}}_{\infty} \, \bigg\lvert  \sum_{j = 1}^{N} \sum_{t = 1}^{T} \big(\sum_{i = 1}^{N} u_{ijt}\big)^2 \bigg\rvert + \\
    & \, \frac{3}{T} \, \norm{(\mu^{\langle 1 \rangle})^{- 1}}_{\infty} \, \bigg\lvert  \sum_{i = 1}^{N} \sum_{j = 1}^{N} \big(\sum_{t = 1}^{T} u_{ijt}\big)^2 \bigg\rvert \\
    =& \, \mathcal{O}_{P}(NT) \, ,
\end{align*}
where we used that $\norm{(\mu^{\langle 1 \rangle})^{- 1}}_{\infty} = \mathcal{O}_{P}(1)$ by Assumption 1 iii) and that
\begin{align*}
    \EX{\CEX{\sum_{i = 1}^{N} \sum_{t = 1}^{T} \big(\sum_{j = 1}^{N} u_{ijt}\big)^2}} =& \, \sum_{i = 1}^{N} \sum_{j = 1}^{N} \sum_{t = 1}^{T} \EX{(u_{ijt})^2} \\
    =& \, \mathcal{O}((NT)^{3 / 2}) \, ,
\end{align*}
by conditional independence and the fact that $\norm{u}_{\infty} \leq 1$. The latter implies that $\sum_{i = 1}^{N} \sum_{t = 1}^{T} (\sum_{j = 1}^{N} u_{ijt})^2 = \mathcal{O}_{P}((NT)^{3 / 2})$. The other bounds, $\sum_{j = 1}^{N} \sum_{t = 1}^{T} (\sum_{i = 1}^{N} u_{ijt})^2 = \mathcal{O}_{P}((NT)^{3 / 2})$ and $\sum_{i = 1}^{N} \sum_{j = 1}^{N} (\sum_{t = 1}^{T} u_{ijt})^2 = \mathcal{O}_{P}((NT)^{3 / 2})$, follow analogously. Bringing all components together and using that $(\mathbb{Q}_{\phi} u)^{\prime} \mathbb{Q}_{\phi} u = \mathcal{O}_{P}(NT)$ and $\norm{\mathbb{Q}_{R} u}_{\infty} = \mathcal{O}_{P}((NT)^{- 3 / 4})$ by Lemma \ref{lemma:linear_operator} ii), we get $(\mathbb{Q} u)^{\prime} \mathbb{Q} u = \mathcal{O}_{P}(NT)$.

For ii), using Loeve's inequality, we have
\begin{align*}
    \big\lvert (\mathbb{Q} u)^{\prime} \diag(\mathbb{Q} u) \mathbb{Q} u \big\rvert =& \, \Big\lvert \sum_{i = 1}^{N} \sum_{j = 1}^{N} \sum_{t = 1}^{T} ((\mathbb{Q} u)_{ijt})^{3} \Big\rvert \\
    \leq& \, 4 \Big\lvert  \sum_{i = 1}^{N} \sum_{j = 1}^{N} \sum_{t = 1}^{T} ((\mathbb{Q}_{\phi} u)_{ijt})^{3} \Big\rvert + 4 \Big\lvert \sum_{i = 1}^{N} \sum_{j = 1}^{N} \sum_{t = 1}^{T} ((\mathbb{Q}_{R} u)_{ijt})^{3} \Big\rvert\\
    \leq& \, 4 \big\lvert (\mathbb{Q}_{\phi} u)^{\prime} \diag(\mathbb{Q}_{\phi} u) \mathbb{Q}_{\phi} u \big\rvert + 4 N^2 T \, \norm{\mathbb{Q}_{R} u}_{\infty}^{3} \, .
\end{align*}
Further, since $\mathbb{Q}_{\phi} = \mathbb{Q}_{\alpha} + \mathbb{Q}_{\gamma} + \mathbb{Q}_{\rho}$, using again Loeve's inequality, it follows that
\begin{align*}
    \big\lvert (\mathbb{Q}_{\phi} u)^{\prime} \diag(\mathbb{Q}_{\phi} u) \mathbb{Q}_{\phi} u \big\rvert \leq& \, 9 \big\lvert (\mathbb{Q}_{\alpha} u)^{\prime} \diag(\mathbb{Q}_{\alpha} u) \mathbb{Q}_{\alpha} u \big\rvert + 9 \big\lvert (\mathbb{Q}_{\gamma} u)^{\prime} \diag(\mathbb{Q}_{\gamma} u) \mathbb{Q}_{\gamma} u \big\rvert + \\
    & \, 9 \big\lvert (\mathbb{Q}_{\rho} u)^{\prime} \diag(\mathbb{Q}_{\rho} u) \mathbb{Q}_{\rho} u \big\rvert \\
    =& \, 9 N \bigg\lvert  \sum_{i = 1}^{N} \sum_{t = 1}^{T} \frac{(\sum_{j = 1}^{N} u_{ijt})^3}{(\sum_{j = 1}^{N} \mu^{\langle 1 \rangle}_{ijt})^3} \bigg\rvert + 9 N \bigg\lvert  \sum_{j = 1}^{N} \sum_{t = 1}^{T} \frac{(\sum_{i = 1}^{N} u_{ijt})^3}{(\sum_{i = 1}^{N} \mu^{\langle 1 \rangle}_{ijt})^3} \bigg\rvert + \\
    & \, 9 T \bigg\lvert  \sum_{i = 1}^{N} \sum_{j = 1}^{N} \frac{(\sum_{t = 1}^{T} u_{ijt})^3}{(\sum_{t = 1}^{T} \mu^{\langle 1 \rangle}_{ijt})^3} \bigg\rvert \\
    \leq& \, \frac{9}{N^2} \, \norm{(\mu^{\langle 1 \rangle})^{- 1}}_{\infty} \, \bigg\lvert  \sum_{i = 1}^{N} \sum_{t = 1}^{T} \big(\sum_{j = 1}^{N} u_{ijt}\big)^3 \bigg\rvert + \\
    & \, \frac{9}{N^2} \, \norm{(\mu^{\langle 1 \rangle})^{- 1}}_{\infty} \, \bigg\lvert  \sum_{j = 1}^{N} \sum_{t = 1}^{T} \big(\sum_{i = 1}^{N} u_{ijt}\big)^3 \bigg\rvert + \\
    & \, \frac{9}{T^2} \, \norm{(\mu^{\langle 1 \rangle})^{- 1}}_{\infty} \, \bigg\lvert  \sum_{i = 1}^{N} \sum_{j = 1}^{N} \big(\sum_{t = 1}^{T} u_{ijt}\big)^3 \bigg\rvert \\
    =& \, \mathcal{O}_{P}(\sqrt{NT}) \, ,
\end{align*}
where we used that $\norm{(\mu^{\langle 1 \rangle})^{- 1}}_{\infty} = \mathcal{O}_{P}(1)$ by Assumption 1 iii) and that
\begin{align*}
    \EX{\CEX{\Big( \sum_{i = 1}^{N} \sum_{t = 1}^{T} \big(\sum_{j = 1}^{N} u_{ijt}\big)^3 \Big)^{2}}} =& \, \sum_{(i, t) = (1, 1)}^{(N, T)} \sum_{j = 1}^{N} \sum_{j^{\prime} = 1}^{N} \sum_{j^{\prime\prime} = 1}^{N} \EX{(u_{ijt})^2 (u_{ij^{\prime}t})^2 (u_{ij^{\prime\prime}t})^2} + \\
    & \, \sum_{(i, t) = (1, 1)}^{(N, T)} \sum_{(i^{\prime}, t^{\prime}) \neq (i, t)}^{(N, T)} \sum_{j = 1}^{N} \sum_{j^{\prime} = 1}^{N} \EX{(u_{ijt})^3} \EX{(u_{i^{\prime}j^{\prime}t^{\prime}})^3} \\
    =& \, \mathcal{O}((NT)^{3}) \, ,
\end{align*}
by conditional independence and the fact that $\norm{u}_{\infty} \leq 1$. The latter implies that $\sum_{i = 1}^{N} \sum_{t = 1}^{T} (\sum_{j = 1}^{N} u_{ijt})^3 = \mathcal{O}_{P}((NT)^{3 / 2})$. The other bounds, $\sum_{j = 1}^{N} \sum_{t = 1}^{T} (\sum_{i = 1}^{N} u_{ijt})^3 = \mathcal{O}_{P}((NT)^{3 / 2})$ and $\sum_{i = 1}^{N} \sum_{j = 1}^{N} (\sum_{t = 1}^{T} u_{ijt})^3 = \mathcal{O}_{P}((NT)^{3 / 2})$, follow analogously. Bringing all components together and using that $(\mathbb{Q}_{\phi} u)^{\prime} \diag(\mathbb{Q}_{\phi} u) \mathbb{Q}_{\phi} u = \mathcal{O}_{P}(\sqrt{NT})$ and $\norm{\mathbb{Q}_{R} u}_{\infty} = \mathcal{O}_{P}((NT)^{- 3 / 4})$ by Lemma \ref{lemma:linear_operator} ii), we get $(\mathbb{Q} u)^{\prime} \diag(\mathbb{Q} u) \mathbb{Q} u = \mathcal{O}_{P}(\sqrt{NT})$.\hfill\qedsymbol

\subsection{Additional technical lemmas}

\begin{lemma}[Central limit theorem]\label{lemma:clt}
    Let Assumption 1 hold. Then,
    \begin{equation*}
        U^{(0)} = (M x)^{\prime} u / (N \sqrt{T})  \overset{d}{\rightarrow} \mathcal{N}(0, \overline{W}) \, ,
    \end{equation*}
    where $\overline{W} = \EX{W}$. 
\end{lemma}

\noindent\textbf{Proof of Lemma \ref{lemma:clt}.} We use Liapunov's central limit theorem, see, for example, \textcite{h2022} Theorem 6.5, to derive the asymptotic distribution of $U^{(0)}$. Let $z_{n,r} = \tilde{x}_{n,r} u_{n,r} = (\tilde{x}_{n,r,1} u_{n,r}, \ldots, \tilde{x}_{n,r, K} u_{n,r})$, and $V$ be a covariance matrix, where $\tilde{x}_{n} = M x$ is a $n \times K$ matrix, $u_{n} = y_{n} - \mu_{n}$ and $\mu_{n} = \mu(x_{n} \beta^{0} + w_{n} \phi^{0})$ are $n$-dimensional vectors, and $n = N^{2}T$. For
\begin{equation*}
    U^{(0)} = \frac{1}{\sqrt{n}} \sum_{r = 1}^{n} z_{n,r} \overset{d}{\rightarrow} \N(0, V)
\end{equation*}
to hold, we need to verify that
\begin{enumerate}[i)]
    \item $\EX{z_{n,r}} = 0$, 
    \item $\frac{1}{n} \sum_{r = 1}^{n} \EX{z_{n,r} (z_{n,r})^{\prime}} \rightarrow V > 0$, and
    \item $\max_{n,r} \EX{\norm{z_{n,r}}_{2}^{2 + \delta}} < \infty$ for some $\delta > 0$.
\end{enumerate}

To verify condition i), note that
\begin{equation*}
    \EX{z_{n,r}} = \EX{\CEX{z_{n,r}}} =  \EX{\tilde{x}_{n,r} (\CEX{y_{n,r}} - \mu_{n, r})} = 0 \, ,
\end{equation*}
since $\CEX{y_{n,r}} = \mu_{n, r}$ is implied by Assumption 1 ii). To verify condition ii), note that
\begin{align*}
     \frac{1}{n} \sum_{r = 1}^{n} \EX{z_{n,r} (z_{n,r})^{\prime}} =& \, \frac{1}{n} \sum_{r = 1}^{n} \EX{\CEX{z_{n,r} (z_{n,r})^{\prime}}} \\
     =& \, \frac{1}{n} \sum_{r = 1}^{n} \EX{\CEX{(u_{n,r})^2} \tilde{x}_{n,r} (\tilde{x}_{n,r})^{\prime}} \\
     =& \, \frac{1}{n} \sum_{r = 1}^{n} \EX{\mu_{n,r}^{\prime} \tilde{x}_{n,r} (\tilde{x}_{n,r})^{\prime}} \\
     =& \, \EX{W} \, ,
\end{align*}
where the last equality follows from $\CEX{(u_{n,r})^2} = \mu_{n,r} (1 - \mu_{n,r}) = \mu_{n,r}^{\prime}$ implied by Assumption 1 ii). It follows that $V = \EX{W} = \overline{W}$. To verify condition iii) (Liapunov condition), we choose $\delta = 1$, and get
\begin{align*}
    \max_{n,r} \EX{\norm{z_{n,r}}_{2}^{3}} =& \, \max_{n,r} \EX{\norm{\tilde{x}_{n,r} u_{n,r}}_{2}^{3}} \\
    \leq& \, \max_{n,r} \big(\EX{(u_{n,r})^{6}}\big)^{1 / 2} \max_{n,r} \big(\EX{\norm{\tilde{x}_{n,r}}_{2}^{6}}\big)^{1 / 2} < \infty \, ,
\end{align*}
where the last inequality follows immediately from the fact that $\norm{u_{n,r}}_{\infty} \leq 1$ and our assumption that $x$ is uniformly bounded, Assumption 1 iii).\hfill\qedsymbol

\begin{lemma}[Normalized profile Hessian]\label{lemma:profile_hessian}
    Let Assumption 1 hold. Then,
    \begin{equation*}
        \lambda_{\min}(W(\beta, \phi)) \geq c_{\min} \, c_{2} > 0 \, ,
    \end{equation*}
    where $\lambda_{\min}(\cdot)$ is the smallest eigenvalue of a matrix. Moreover, $\norm{(W(\beta, \phi))^{- 1}}_{2} = \mathcal{O}_{P}(1)$.
\end{lemma}

\noindent\textbf{Proof of Lemma \ref{lemma:profile_hessian}.} Our proof closely follows the proof strategy of Lemma 4 in \textcite{cfw2020}. Let 
\begin{align*}
    W(\beta, \phi) =& \, \frac{1}{N^{2}T} (M(\beta, \phi) x)^{\prime} \diag(\mu^{\langle 1 \rangle}(x \beta + w \phi)) x \\
    =& \, \frac{1}{N^{2}T} (M(\beta, \phi) x)^{\prime} \diag(\mu^{\langle 1 \rangle}(x \beta + w \phi)) M(\beta, \phi) x  \\
    =& \, \frac{1}{N^{2}T} (\tilde{x}(\beta, \phi))^{\prime} \diag(\mu^{\langle 1 \rangle}(x \beta + w \phi)) \tilde{x}(\beta, \phi)
\end{align*}
denote the normalized profile Hessian defined in \eqref{eq:profile_hessian}, where we used that $M(\beta, \phi)$ is a projection matrix and that $\tilde{x}(\beta, \phi) = M(\beta, \phi) x$. Then, by the Courant–Fischer–Weyl min-max principle,
\begin{align*}
    \lambda_{\min}(W(\beta, \phi)) =& \, \underset{\{\Delta \in \mathbb{R}^{K} \colon \norm{\Delta} = 1\}}{\min} \; \frac{1}{N^2T} \Delta^{\prime} ((\tilde{x}(\beta, \phi))^{\prime} \diag(\mu^{\langle 1 \rangle}(x \beta + w \phi)) \tilde{x}(\beta, \phi)) \Delta \\
    =& \, \underset{\{\Delta \in \mathbb{R}^{K} \colon \norm{\Delta} = 1\}}{\min} \; \frac{1}{N^2T} \sum_{i = 1}^{N} \sum_{j = 1}^{N} \sum_{t = 1}^{T} \mu_{ijt}^{\langle 1 \rangle}(\beta, \phi) (\tilde{x}_{ijt}(\beta, \phi) \Delta)^{2} \\
    \geq& \, \underset{\{\Delta \in \mathbb{R}^{K} \colon \norm{\Delta} = 1\}}{\min} \; \frac{c_{\min}}{N^2T} \sum_{i = 1}^{N} \sum_{j = 1}^{N} \sum_{t = 1}^{T} (x_{ijt} \Delta -  w_{ijt} \hat{\pi})^{2} \\
    =& \, \underset{\{\Delta \in \mathbb{R}^{K} \colon \norm{\Delta} = 1\}}{\min} \; \underset{\{\pi \in \mathbb{R}^{2NT + N^2}\}}{\min} \;  \frac{c_{\min}}{N^2T} \sum_{i = 1}^{N} \sum_{j = 1}^{N} \sum_{t = 1}^{T} (x_{ijt} \Delta - w_{ijt} \pi)^{2} \, ,
\end{align*}
where we used that, by Assumption 1 iii) and iv), $0 < c_{\min} \leq \mu_{ijt}^{\langle 1 \rangle}(\beta, \phi) \leq c_{\max} < \infty$ and
\begin{equation*}
    \underset{\{\Delta \in \mathbb{R}^{K} \colon \norm{\Delta} = 1\}}{\min} \; \underset{\{\pi \in \mathbb{R}^{2NT + N^2}\}}{\min} \;  \frac{1}{N^2T} \sum_{i = 1}^{N} \sum_{j = 1}^{N} \sum_{t = 1}^{T} (x_{ijt} \Delta - w_{ijt} \pi)^{2} \geq c_{2} > 0 \, ,
\end{equation*}
respectively. Bringing all components together, we conclude that $\lambda_{\min}(W(\beta, \phi)) \geq c_{\min} \, c_{2} > 0$, which implies $\norm{(W(\beta, \phi))^{- 1}}_{2} = \mathcal{O}_{P}(1)$. \hfill\qedsymbol

\begin{lemma}[Score of incidental parameters]\label{lemma:ip_score}
    Let Assumption 1 hold. Then,
    \begin{equation*}
        \bignorm{(w^{\prime} u) / \sqrt{NT}}_{\infty} = \mathcal{O}_{P}((NT)^{- 1 / 4}) \, .
    \end{equation*}
\end{lemma}

\noindent\textbf{Proof of Lemma \ref{lemma:ip_score}.} For each $g \in \{1, \ldots, 2NT + N^2\}$, we have
\begin{align*}
    \EX{\CEX{((NT)^{- 1 / 4} e_{g}^{\prime} w^{\prime} u)^2}} =& \, \frac{1}{\sqrt{NT}} \, \EX{\CEX{\Big(\sum_{i = 1}^{N} \sum_{j = 1}^{N} \sum_{t = 1}^{T} w_{ijt} e_{g} u_{ijt}\Big)^{2}}} \\
    =& \, \frac{1}{\sqrt{NT}} \, \sum_{i = 1}^{N} \sum_{j = 1}^{N} \sum_{t = 1}^{T} w_{ijt} e_{g} \, \EX{(u_{ijt})^{2}} \\
    =& \, \mathcal{O}(1) \, ,
\end{align*}
by conditional independence, the sparsity of $w$, i.e.\ $\norm{w}_{1} = \max(N, T)$, and the fact that $\norm{u}_{\infty} \leq 1$. Since $\EX{e_{g}^{\prime} w^{\prime} u} = 0$ for all $g \in \{1, \ldots, 2NT + N^2\}$, Hoeffding's Lemma (see \textcite{h1963}) implies that $\{(NT)^{- 1 / 4} e_{g}^{\prime} w^{\prime} u\}_{g = 1}^{2NT + N^2}$ is sub-Gaussian with variance proxy $\sigma_{g}^{2} < \infty$. Thus, we can use an upper bound for sub-Gaussian maxima, see, for example, \textcite{w2019a},
\begin{equation*}
    \EX{\underset{\{g \in \{1, \ldots, 2NT + N^2\}\}}{\max} \big\lvert (NT)^{- 1 / 4} e_{g}^{\prime} w^{\prime} u \big\rvert} \leq C \sqrt{\log(NT)} \, ,
\end{equation*}
where $C < \infty$ is a universal constant independent of the sample size. Bringing all components together we conclude that $\norm{(w^{\prime} u) / \sqrt{NT}}_{\infty} = \mathcal{O}_{P}((NT)^{- 1 / 4})$. \hfill\qedsymbol

\begin{lemma}[Convergence rates]\label{lemma:convergence_rates}
    Let Assumption 1 hold. Then,
    \begin{enumerate}[i)]
        \item $\norm{\hat{\beta} - \beta^{0}}_{2} = \mathcal{O}_{P}((NT)^{- 3 / 4})$,
        \item $\norm{\hat{\phi}(\hat{\beta}) - \phi^{0}}_{\infty} = \mathcal{O}_{P}((NT)^{- 1 / 4})$.
    \end{enumerate}
\end{lemma}

\noindent\textbf{Proof of Lemma \ref{lemma:convergence_rates}.} For i), by Lemma \ref{lemma:first_order_expansions} i), we have
\begin{equation}
\label{eq:score_expansion_first_order}
    \frac{1}{\sqrt{NT}} (\check{M} x)^{\prime} \diag(\check{\mu}^{\langle 1 \rangle}) x (\hat{\beta} - \beta^{0}) = \frac{1}{\sqrt{NT}} (\check{M} x)^{\prime} u\, . 
\end{equation}
Re-arranging \eqref{eq:score_expansion_first_order} yields
\begin{equation*}
    N\sqrt{T} (\hat{\beta} - \beta^{0}) = \check{W}^{- 1} \frac{1}{N \sqrt{T}} (\check{M} x)^{\prime} u  \, ,
\end{equation*}
where $\check{W}$ is the normalized profile Hessian evaluated at $\check{\beta}$ and $\hat{\phi}(\check{\beta})$. It follows that
\begin{equation*}
    N\sqrt{T} \norm{\hat{\beta} - \beta^{0}}_{2} \leq K \, \norm{\check{M} x}_{\max} \, \norm{\check{W}^{- 1}}_{2} \, \frac{1}{N \sqrt{T}} \, \norm{u}_{2} \, ,
\end{equation*}
where $\norm{(\check{W})^{- 1}}_{2} = \mathcal{O}_{P}(1)$ by Lemma \ref{lemma:profile_hessian} and $\norm{\check{M} x}_{\max} = \mathcal{O}_{P}(1)$ by Assumption 1 iii). Further,
\begin{align*}
    \EX{\CEX{\norm{u}_{2}^{2}}} =& \, \sum_{i = 1}^{N} \sum_{j = 1}^{N} \sum_{t = 1}^{T} \sum_{i^{\prime} = 1}^{N} \sum_{j^{\prime} = 1}^{N} \sum_{t^{\prime} = 1}^{T} \EX{\CEX{u_{ijt} u_{i^{\prime}j^{\prime}t^{\prime}}}} \\
    =& \, \sum_{i = 1}^{N} \sum_{j = 1}^{N} \sum_{t = 1}^{T} \EX{u_{ijt}^{2}} \\
    =& \, O(N^{2} T) \, ,
\end{align*}
by conditional independence and the fact that $\norm{u}_{\infty} \leq 1$. Thus, $\norm{u}_{2} = \mathcal{O}_{P}(N\sqrt{T})$. Bringing all components together we conclude that $\norm{\hat{\beta} - \beta^{0}}_{2} = \mathcal{O}_{P}((NT)^{- 3 / 4})$.

For ii), by Lemma \ref{lemma:first_order_expansions} ii), we have 
\begin{equation*}
    \hat{\phi}(\hat{\beta}) - \phi^{0} = - \frac{1}{\sqrt{NT}}  \check{H}^{- 1} w^{\prime} \diag(\check{\mu}^{\langle 1 \rangle}) x (\hat{\beta} - \beta^{0}) + \frac{1}{\sqrt{NT}} \check{H}^{- 1} w^{\prime} u \, .
\end{equation*}
Further, by the triangle inequality and the Hoelder's inequality, it follows that
\begin{align*}
    \norm{\hat{\phi}(\hat{\beta}) - \phi^{0}}_{\infty} \leq& \, \underset{\{g \in \{1, \ldots, 2 N T + N^{2}\}\}}{\max} \big\lvert e_{g}^{\prime} \check{H}^{- 1} w^{\prime} \diag(\check{\mu}^{\langle 1 \rangle}) x (\hat{\beta} - \beta^{0}) \big\rvert / \sqrt{NT} + \\
    & \underset{\{g \in \{1, \ldots, 2 N T + N^{2}\}\}}{\max} \big\lvert e_{g}^{\prime} \check{H}^{- 1} w^{\prime} u \big\rvert / \sqrt{NT} \\
    \leq& \, \underset{\{g \in \{1, \ldots, 2 N T + N^{2}\}\}}{\max} \big\lvert e_{g}^{\prime} w^{\prime} x (\hat{\beta} - \beta^{0}) \big\rvert \, \norm{\check{\mu}^{\langle 1 \rangle}}_{\infty} \norm{\check{H}^{- 1}}_{\max} / \sqrt{NT} + \\
    & \, \underset{\{g \in \{1, \ldots, 2 N T + N^{2}\}\}}{\max} \big\lvert e_{g}^{\prime} w^{\prime} u \big\rvert \, \norm{\check{H}^{- 1}}_{\max} / \sqrt{NT} \\
    \leq& \, \underset{\{g \in \{1, \ldots, 2 N T + N^{2}\}\}}{\max} \norm{e_{g}}_{1} \, \norm{w^{\prime} x (\hat{\beta} - \beta^{0})}_{\infty} \, \norm{\check{\mu}^{\langle 1 \rangle}}_{\infty} \norm{\check{H}^{- 1}}_{\max} / \sqrt{NT} + \\
    & \, \underset{\{g \in \{1, \ldots, 2 N T + N^{2}\}\}}{\max} \norm{e_{g}}_{1} \, \norm{w^{\prime} u}_{\infty} \norm{\check{H}^{- 1}}_{\max} / \sqrt{NT} \\
    \leq& \, \norm{w^{\prime} x}_{\infty} \, \norm{\hat{\beta} - \beta^{0}}_{2} \, \norm{\check{\mu}^{\langle 1 \rangle}}_{\infty} \norm{\check{H}^{- 1}}_{\max} / \sqrt{NT} + \\
    & \, \norm{w^{\prime} u}_{\infty} \, \norm{\check{H}^{- 1}}_{\max} / \sqrt{NT} \\
    =& \, \mathcal{O}_{P}((NT)^{- 1 / 4}) \, ,
\end{align*}
where $\norm{w^{\prime} x}_{\infty} = \mathcal{O}_{P}((NT)^{1 / 2})$ by Assumption 1 iii), $\norm{\hat{\beta} - \beta^{0}}_{2} = \mathcal{O}_{P}((NT)^{- 3 / 4})$ shown above, $\norm{\check{\mu}}_{\infty} = \mathcal{O}_{P}(1)$ by Assumption 1 iii), $\norm{w^{\prime} u}_{\infty} = \mathcal{O}_{P}((NT)^{1 / 4})$ by Lemma \ref{lemma:ip_score}, and
\begin{align*}
    \norm{\check{H}^{- 1}}_{\max} \leq& \, \norm{\check{D}^{- 1}}_{\max} + \norm{\check{H}^{- 1} - \check{D}^{- 1}}_{\max} \\
    \leq& \, \norm{(\check{\mu})^{- 1}}_{\infty} + \norm{\check{H}^{- 1} - \check{D}^{- 1}}_{\max} \\
    =& \, \mathcal{O}_{P}(1)
\end{align*}
by Assumption 1 iii) and Lemma \ref{lemma:inverse_ip_hessian}. \hfill\qedsymbol

\clearpage
\printbibliography
\end{document}